\documentclass[12pt, preprint]{aastex}
\usepackage{graphicx}

\def\msun{M$_\odot$}
\def\msunyr{M$_\odot/{\rm yr}\ $}

\begin{document}

\title{Outflow Feedback Regulated Massive Star Formation \\
in Parsec-Scale Cluster Forming Clumps}

\author{Peng Wang$\!$\altaffilmark{1} , Zhi-Yun  
Li$\!$\altaffilmark{2}, Tom Abel$\!$\altaffilmark{1} and Fumitaka
  Nakamura$\!$\altaffilmark{3} }
\altaffiltext{1}{KIPAC, SLAC, and Physics Department, Stanford
  University, Menlo Park, CA 94025} 
\altaffiltext{2}{Astronomy Department, P.O. Box 400325, 
University of Virginia, Charlottesville, VA 22904; zl4h@virginia.edu}
\altaffiltext{3} {Faculty of Education, Niigata University, 
8050 Ikarashi-2, Niigata 950-2181, Japan; also visiting astronomer 
at Institute of Space and Astronautical Science, Japan Aerospace 
Exploration Agency, 3-1-1 Yoshinodai, Sagamihara, Kanagawa 
229-8510, Japan}

\begin{abstract}
  We investigate massive star formation in turbulent, magnetized, parsec-scale
  clumps of molecular clouds including protostellar outflow feedback using 
  three dimensional numerical simulations of effective resolution $2048^3$. 
  The calculations are carried out using a block structured adaptive mesh 
  refinement code that solves the ideal MHD equations including self-gravity 
  and implements accreting sink particles. We find that, in the absence of 
  regulation by magnetic fields and outflow feedback, massive stars form  
  readily in a turbulent, moderately condensed clump of $\sim 1,600$~\msun 
  (containing $\sim 10^2$ initial Jeans masses), along with a cluster
  of hundreds of lower mass stars. The massive stars are fed at high 
  rates by (1) transient 
  dense filaments produced by large-scale turbulent compression at  early 
  times,  and (2) by the clump-wide global collapse resulting from turbulence 
  decay at late times. In both cases, the bulk of the massive star's mass is 
  supplied from outside a 0.1~pc-sized ``core" that surrounds the star. In 
  our simulation, the massive star is clump-fed rather than core-fed.  The 
  need for large-scale feeding makes the massive star formation prone to 
  regulation by outflow feedback, which directly opposes the feeding 
  processes. 
  The outflows reduce the mass accretion rates onto the massive stars by 
  breaking up the dense filaments that feed the massive star formation at 
  early times, and by collectively 
  slowing down the global collapse that fuel the 
  massive star formation at late times. The latter is aided 
  by a moderate magnetic field of strength in the observed range 
  (corresponding to a dimensionless clump mass-to-flux ratio $\lambda\ 
  \sim$ a few); the field allows the outflow momenta to be deposited 
  more efficiently inside the clump.  We conclude that the massive 
  star formation in our simulated 
  turbulent, magnetized, parsec-scale clump is outflow-regulated and 
  clump-fed (ORCF for short). An important implication is that the 
  formation of low-mass stars in a dense clump can affect the 
  formation of massive stars in the same clump, through their 
  outflow feedback on the clump dynamics. In a companion paper, 
  we discuss the properties of the lower mass cluster members 
  formed along with the massive stars, including their mass 
  distribution and spatial clustering.  
\end{abstract}
\maketitle

\section{Introduction}

The most important factor that determines whether a massive star can
form or not is its mass accretion rate (Larson \& Starrfield 1971; for
recent reviews, see McKee \& Ostriker 2007; Zinnecker \& Yorke
2007). It follows from the fact that the outward radiative force on
the accretion flow onto a forming massive star increases rapidly with
the stellar mass. If the mass accretion rate is too low, there would
be insufficient ram pressure to overcome the radiation pressure near
the dust sublimation radius, and the infall would be choked. In the
spherical model of Wolfire \& Cassinelli (1987), the required minimum
accretion rate ranges from $\sim 10^{-4}$ to $\sim 10^{-3}$~\msunyr
for stars in the mass range $30 - 100$~\msun, although the requirement
can be relaxed somewhat if the accretion flow is flattened, by, e.g.,
magnetic support (Nakano 1989) or rotation (Jijina \& Adams 1996).  In
addition to overcoming the radiation pressure, a high accretion rate
can also greatly modify the internal structure and appearance of a
growing massive star (e.g., Hosokawa \& Omukai 2009). The associated
high accretion luminosity may help heat up the region surrounding the
protostar, perhaps suppressing the potential fragmentation that
may adversely affect massive star formation at the early stages
(Krumholz \& McKee 2008).

A high mass accretion rate may not be difficult to achieve 
in principle, given that most, if not all, massive stars 
are thought to form in dense, massive, cluster-forming 
clumps (McKee \& Ostriker 2007; Zinnecker \& Yorke 2007). 
For example, a parsec-sized dense clump of $10^3$\msun 
would have an average density of $6.8\times10^{-20}$~g~cm$^{-3}$,
corresponding to a global free fall time $t_{\rm c, ff}= 
0.26$~Myr. The ratio of the clump mass and free-fall time yields 
a characteristic accretion rate ${\dot M}_{\rm c,ff} = 3.9 
\times 10^{-3}$~\msunyr (e.g., Cesaroni 2005) which, if 
channeled to a single object, would have a sufficiently 
large ram pressure to overcome the radiation pressure from 
even the most massive stars. In practice, the mass accretion 
rate onto any individual star will be reduced relative to 
the characteristic free-fall rate by several factors, 
such as the turbulence, magnetic fields, protostellar 
outflows, and radiative feedback. These factors are included 
in our simulations, except the last one. 

Supersonic turbulence is observed ubiquitously in dense 
clumps of star cluster and massive star formation (e.g., 
Garay 2005). The turbulence is expected to reduce the 
mass accretion rate onto any individual star relative 
to ${\dot M}_{\rm c,ff}$ in at least two ways. First, 
it resists the global collapse, thereby reducing the total 
amount of gas that collapses into stars per unit time.
Second, it fragments the clump material into many 
centers of collapse, creating a cluster of stars each 
accreting at a fraction of the total rate. These effects 
of the turbulence are difficult to quantify analytically; 
numerical simulations are required (Mac Low \& Klessen 
2004). For example, using SPH simulation with 
sink particles, Bonnell, Bate \& Vine (2003) found that 
a dense turbulent clump of $10^3$\msun in mass and 0.5 pc 
in radius (corresponding to ${\dot M}_{\rm c,ff}=
5.4\times 10^{-3}$\msunyr) produced a cluster of about
400 stars. In their simulation, the most massive star 
(of 27~\msun) formed from an average accretion rate of order 
$\sim 10^{-4}$\msunyr, much smaller than the global
free-fall rate. In this paper, we will explore the 
effects of the turbulence on the stellar mass accretion 
further with a grid-based code, Enzo, including adaptive 
mesh refinement (AMR) and sink particles. The grid-based 
code also enables us to quantify, in addition, the 
effects of magnetic fields, which are generally 
difficult to treat in SPH (see, however, Price \& Bate 
2008). 

There is ample observational evidence for magnetic fields in regions
of massive star formation. In the nearest region of active massive
star formation, OMC1, the well-ordered polarization vectors of
submillimeter dust continuum leave little doubt that the (slightly
pinched) magnetic field is dynamically important (e.g., Vaillancourt
et al. 2008). A CN Zeeman measurement yields a line-of-sight field
strength of 0.36~mG, corresponding to a magnetic energy density that
is higher than the turbulent energy density and a mass-to-flux ratio
(before geometric correction) that is 4.5 times higher than the
critical value $(2\pi G^{1/2})^{-1}$ (Falgarone et al. 2008). The
mass-to-flux ratio is close to the median value inferred by Falgarone
et al. for a sample of a dozen or so high-mass star forming regions
where the CN Zeeman measurements are made. Their best estimate for the
mean dimensionless mass-to-flux ratio (in units of the critical value)
is $\lambda \sim 2$ after geometric corrections, although it can be
uncertain by a factor of 2 in either direction. A goal of our
calculations is to determine how a magnetic field of this magnitude
affects the stellar mass accretion rate, using an MHD version of the
Enzo code (Wang \& Abel 2009).

The third element that we consider in this paper are protostellar outflows, 
which are routinely observed around forming stars of both low and high
masses (see the review by Richer et al. 2000).  Their effects are
expected to be particularly important in the dense clumps of active
cluster formation, where many stars are formed close together in both
space and time (Li \& Nakamura 2006; Norman \& Silk 1980).  A case in
point is the dense clump associated with the reflection nebula
NGC~1333 in the Perseus molecular cloud, which is sculpted by dozens of
outflows detectable in CO, optical forbidden lines and H$_2$ emission
(Sandell \& Knee 2001; Walawender et al. 2008; Maret et al. 2009). The
outflows appear to inject enough momentum into the dense clump to
replenish the turbulence dissipated in this region (see, however, Maury 
et al. 2009 for a discussion of NGC 2264, where the current generation 
of active outflows appear incapable of supporting the cluster-forming 
clump). It is possible that a good fraction, perhaps the majority, of the 
cluster members are formed during the time when the clump turbulence
is driven by protostellar outflows  (Nakamura \& Li 2007, 
Matzner 2007, Carroll et al.  2008, Swift \& Welch 2008). We will improve 
on the previous outflow feedback simulations of Nakamura \& Li (2007) 
by including sink particles, making the outflow injection more
continuous, and dramatically increasing the numerical 
resolution. These improvements  enable us to evaluate the effects of 
the outflows on the accretion rates onto individual stars, especially 
massive stars, which are the focus of this paper. We will describe 
the properties of the lower mass cluster members formed in our 
simulations, including their mass distribution and spatial 
clustering, in a companion paper. The effects of radiative 
feedback, ignored in our current calculations, are discussed in 
\S~\ref{limitation}.

The rest of the paper is organized as follows. In \S~\ref{method}, we 
describe the numerical method and simulation setup. These are followed 
by two result sections, focusing first on the global star formation history 
and clump dynamics (\S~\ref{global}) and then on the massive stars 
(\S~\ref{massivestars}). We find that the massive stars in our simulations 
are formed through neither the collapse of pre-existing 0.1~pc-sized 
turbulent cores nor competitive accretion; their formation is controlled 
to a large extent by the clump dynamics, which are found to be regulated 
strongly by the collective outflow feedback from all accreting 
stars and, to a lesser extent, by a moderate magnetic field. In 
\S~\ref{discussion}, we discuss this new scenario 
of ``outflow-regulated clump-fed" (ORCF for short) massive star 
formation and contrast it with the existing scenarios. The last 
section contains a brief summary.

\section{Numerical Method and Simulation Setup}
\label{method}

\subsection{Magnetohydrodynamics}

Our simulations are performed using the parallel AMR code, Enzo, with a
newly added solver for unsplit conservative hydrodynamics and
magnetohydrodynamics (Wang \& Abel 2009). As usual, at the heart of 
the MHD solver lies the enforcement of the divergence-free condition 
$\nabla\cdot B=0$. There are three classes of method for this
purpose: constraint transport (CT, Evans \& Hawley 1988),
projection method and hyperbolic clean (see, e.g., T{\'o}th 2000 
for a review). We have evaluated all these methods and found that
divergence cleaning is most suitable for the problem at hand, which 
involves rather extreme variation in physical quantities (including
the field strength) due to both gravitational collapse and protostellar 
outflows. The fast and robust method of hyperbolic clean of  
Dedner et al. (2002) is adopted. It enables us to follow the formation
of hundreds of stars over several dynamical times. We monitor the 
value of $\nabla\cdot B$ to ensure that it is small in all of our 
MHD simulations.

\subsection{Sink Particles: Creation, Accretion and Merging}

Truelove et al. (1997) pointed out that in order to avoid artificial
fragmentation, one need to use at least four cells to resolve the
Jeans length. For a given cell size at the finest level of refinement, 
the condition translates into a maximum density, which we will call 
the ``Jeans density'', above which artificial fragmentation may 
happen. Since it is difficult in
large-scale simulations such as ours to resolve the Jeans length all
the way to the stellar density, sink particles are used to handle the
gravitational collapse beyond the Jeans density. This approach has
previously been successfully used in both SPH (e.g., Bate et al. 1995) 
and grid-based simulations of star formation (e.g., Krumholz et al. 2004;
Offner et al. 2009).

Our procedure for implementing the sink particles is as follows. When
a cell violates the Truelove criterion on the highest refinement
level, a sink particle is inserted at the center of that cell. The
initial mass of the sink particle is calculated such that after
subtracting the sink mass, the cell will be at the Jeans density. The
initial velocity of the sink particle is calculated using momentum
conservation.

After creation, the sink particle accretes gas from its host cell 
according to a prescription inspired by the Bondi-Hoyle (BH) formula 
(Ruffert 1994),
\begin{equation}
\dot M_{BH} = 4\pi \rho_\infty r_{BH}^2\sqrt{1.2544c_{\infty}^2
+v_\infty^2}, \label{BH}
\end{equation}
where $r_{BH}$ is the Bondi radius, $\rho_\infty, v_\infty$ and
$c_\infty$ are the gas density, velocity and sound speed of the 
(uniform) medium far from the point mass.

Even though equation (\ref{BH}) is not strictly applicable to the  
highly turbulent medium under investigation, we will follow 
Krumholz et al. (2004) and use it as a guide to construct a 
prescription for sink particle accretion. Since the cloud 
material is assumed isothermal, it is natural to set $c_{\infty}$ 
to the isothermal sound speed $c_s$. For $v_{\infty}$, we adopt
the simple prescription $v_\infty=v_{cell}-v_{sink}$, where $v_{cell}$ 
is the flow velocity in the cell and $v_{sink}$ the velocity of the
sink particle. We compute
the Bondi radius in equation~(\ref{BH}) using $r_{BH}=G M_{sink}
/(c_s^2+v_\infty^2)$ where $M_{sink}$ is the sink mass. If $r_{BH}$ 
is smaller than the cell length $\Delta x$, the cell density 
$\rho_{cell}$ is used for $\rho_\infty$. Otherwise an extrapolation 
assuming an $r^{-1.5}$ density profile is used. In other words, 
we set $\rho_\infty = \rho_{cell}\ {\rm min} [1.0,
 (\Delta x/r_{BH})^{1.5}]$, where $\Delta x=200$~AU is the size
of our finest cells. The amount of gas accreted during a single 
time step $\Delta t$ is then $\dot M_{BH}\Delta t$. The amount 
may or may not match the true value exactly at any given time, 
given the crudeness of the above prescription. This deficiency is 
corrected in a second step, through sink particle merging. 

Sink particle merging is needed not only to ensure the correctness of
the mass accretion algorithm, but also to save computation time when
many sink particles are created. It is a tricky problem, because we
want to eliminate, on the one hand, the artificial particles created by
the mismatch between the true accretion rate and that given by
equation~(\ref{BH}), and to preserve, on the other hand, the
legitimate stellar seeds. It requires sub-grid information that is not
available in the simulation. At this early stage in the development of
the sink particle technique, our design principle of the merging 
recipe is to make
it as simple as possible and with as few parameters as possible. We
have experimented with a number of recipes, and settled on one with
two parameters: a merging mass $M_{merg}$ and a merging distance
$l_{merg}$. Using $M_{merg}$ we divide the sink particles into a
``small'' and a ``big'' group. At every time step, the merging is done
in two steps. First, for each small particle, we search within the
merging distance for its nearest big particle. If the search is
successful, we merge this particle to the found big particle. Second,
for all small particles that remain after the first step, we group
them use a Friend-of-Friend algorithm with the merging distance
$l_{merg}$ as the linking length (Davis et al. 1985). In this work, we
use $M_{merg}=0.01$ \msun and $l_{merg}=5 \Delta x$, which is about
$1000$ AU. Experimentation shows that increasing $M_{merg}$ to $0.1$
\msun or $l_{merg}$ to $10 \Delta x$ does not change the star
formation rate or the stellar mass distribution significantly.
Lowering the merging distance to $3 \Delta x$ has little effect on the
total star formation rate, but can change the mass of a star by up to
$50\%$, probably because the flow pattern within a few cells of a sink
particle is not well resolved. The flow so close to a forming
  star may be strongly affected by radiative heating (Offner et
  al. 2009; Bate 2009), which is not included in our simulations. For
these reasons, we believe that the properties of circumstellar disks 
and binaries may not be reliably captured by our simulation; they 
are not discussed further in the paper.

With the inclusion of sink particle merging, the crude prescription
for sink particle accretion based on the BH formula~(\ref{BH}) does 
not appear to pose any serious problem. We have experimented 
with changing the parameters in the formula and did not see any 
significant difference in either the stellar mass accretion rate 
or the final stellar mass distribution. As stressed by Krumholz 
et al. (2004), the reason is probably that, when the BH formula 
underestimates the true accretion rate, the sink particle creation 
routine would transform the un-accreted gas into small sink 
particles, which would merge quickly with the original 
particle and restore the correct rate. On the other hand, when 
the BH formula overestimates the true rate, which happens rarely 
and only when the sink particle is massive, the accretion flow 
is typically supersonic near the particle, and the over-accretion 
does not affect the flow further away. To avoid numerical instabilities, 
we restrict the maximum amount of the gas accreted in a single time-step 
to be less than $25\%$ of the cell mass.

Finally, the gravity of the sink particles is calculated using the 
standard particle-mesh method, and their positions and velocities 
are updated using the leapfrog method. 

\subsection{Protostellar Outflows}

Protostellar outflows are a crucial element of our model. In
large-scale global simulations such as ours, it is unfortunately
impossible to follow their production from first principles. An
observationally motivated prescription is used instead. We assume that
the protostellar outflow momentum injection rate is proportional to
the stellar mass accretion rate, with a mass dependent proportionality
constant $P_*=P_0 (M_*/M_0)^{1/2}$ where $P_0=16$~km/s and $M_0=1$
\msun. The mass dependence is chosen to reflect the fact that the
outflows from high mass stars tend to be more powerful even after
correcting for their larger masses (Richer et al. 2000, D. Shepherd,
private communication).  We represent the outflow by injecting a
momentum $\Delta P=P_*\Delta M$ into the surrounding gas every time
the sink particle has gained a mass increment $\Delta M$. Ideally, one
would like to make $\Delta M$ as small as possible, to render the
injection continuous. However, we find continuous injection is too
time consuming, because of the small Courant timestep associated with 
the fast outflow speed and small grid size near
the sink particles. Furthermore, if $\Delta M$ is much less than the
mass of an injecting cell, the injected momentum would have little
dynamical effect on the cell. These considerations led us to adopt
$\Delta M=0.1$ \msun. We have experimented with $\Delta M=0.2$ \msun,
and it did not change the results significantly. To avoid too high an
outflow speed when a surrounding cell happens to have a low density,
we take $10\%\Delta M$ out of the sink particle, and divide it evenly
between the injecting cells. For simplicity, we inject the momentum 
in a bipolar jet, with the jet direction parallel (and anti-parallel) 
to the local magnetic field direction of the host cell when the jet is 
first injected (the jet direction is kept fixed in time) and a jet 
width of five finest cells.

\subsection{Initial and Boundary Conditions}
\label{setup}

We consider a dense clump of molecular cloud that has an initial 
density profile with a uniform central region,
\begin{equation}
\rho (r) = {\rho_c\over 1+(r/r_c)^2}, \label{denprofile}
\end{equation}
where $r_c = L/6$ and $L$ is the length of the simulation box.  The
central density $\rho_c=1.0\times10^{-19}$ g cm$^{-3}$ and the box
size is $L=2$ pc. This gives a cloud mass within a sphere of 1 pc in
radius of $M_s=1215$ \msun. Since a square box with $L=2$~pc is used,
the total mass is somewhat larger, with $M_{tot}=1641$ \msun. Our
choice of cloud density, mass and mass distribution is motivated by
observations of infrared dark clouds (e.g., Butler \& Tan 2009), which
are thought to be future sites of cluster and massive star formation;
they tend to be centrally condensed even before the onset of massive
star formation.

The equation of state is isothermal, with a sound speed 
$c_s=0.265$ km/s, which corresponds to a temperature of $T=20$~K 
for a mean molecular weight $\mu=2.33$. 

A cloud with a density profile given by equation~(\ref{denprofile}) has a 
gravitational potential energy ${\rm E_G}=0.7 GM_s^2/R$, which 
yields a virial parameter $\alpha_{vir}=2{\rm E_K}/{\rm E_G}=R 
\sigma^2/(0.7GM_s)$, where $\sigma$ is the 3D rms velocity. In 
terms of the 3D Mach number $\mathcal{M}\equiv \sigma/c_s$, we have
\begin{equation}
\alpha_{vir} = 1.5 \left({\mathcal{M}\over 9}\right) .
\end{equation}
The initial Mach number is set to $\mathcal{M}=9$. Following the
standard treatment in supersonic turbulence simulations (e.g. Mac Low
et al. 1998), we impose a velocity field of power spectrum
$v_k^2\propto k^{-4}$ with a minimum wave number 2 and maximum wave
number 10 in units of inverse box length.

In models with a magnetic field, the field is assumed to be 
initially uniform in the $z$ direction, with a strength given by
\begin{equation}
B_0 = 1.0\times10^{-4} \left({\lambda_{tot}\over 1.4}\right)^{-1} \ 
\ {\rm Gauss},
\end{equation}
where $\lambda_{tot} \equiv 2\pi G^{1/2} M_{tot}/(\pi B_0 L^2/4)=1.4$ 
is the mass-to-flux ratio in units of the critical value. Because 
the adopted density profile is centrally condensed, the central 
part of the clump within 
$r_c$ has a larger mass-to-flux ratio $\lambda_c=2.5\lambda_{tot} 
=3.3$. Thus the envelope is more magnetized than the central part. 
Our adopted degree of magnetization is in the range inferred by
Falgarone et al.~(2008).

For all simulations, the top grid has resolution $128^3$ and the
maximum refinement level is 4, which corresponds to a highest
resolution of $200$ AU. With this resolution and our sink particle
algorithms, the risk of over producing low mass stars is reduced 
in our model even though we do not include radiative heating, 
which happens predominantly on the disk scales (Krumholz et al. 
2007, Bate 2009). We
will not be confident about the mass spectrum at the lowest mass end.
However, the formation of massive stars, the main focus of this paper,
is adequately resolved.

The boundary condition is periodic for all hydro and MHD quantities 
and gravity. In the model with protostellar outflows, to prevent 
the high speed outflows from re-entering the other side of the 
boundary and thus artificially increasing their effects, we 
reduce their speeds by a factor of $10$ when they reach the 
boundary.

We consider four models of increasing complexities, starting with 
a base model (BASE) that is initially neither turbulent nor 
magnetized. We then add to this base model, one by one, an initial 
turbulence (HD), magnetic fields (MHD) and outflows (WIND). The 
models are summarized in Table~1 and discussed in turn below. 
Movies for the HD, MHD and WIND models are included in the
electronic version of the paper.  

\section{Star Formation History and Global Gas Dynamics} 
\label{global}

Our centrally condensed clump described in \S~\ref{setup} contains
many thermal Jeans masses. In the absence of any support in addition
to thermal pressure, the clump collapses dynamically towards the
center, forming an object within 0.2 Myr (see Fig.~1), close to the
free-fall time of the central plateau part of the initial clump
configuration $t_{\rm ff,0}=0.21$~Myr. There is a rapid initial
increase in the mass accretion rate, characteristic of free-fall
collapse of regions of flat density distribution. The accretion rate
settles into a more or less constant value, as the material in the
envelope part of the initial clump (that has an $r^{-2}$ density
distribution) starts to fall into the center. This nearly constant
rate of mass accretion is reminiscent of the collapse of the singular
isothermal sphere (SIS; Shu 1977), although the rate ${\dot M} \sim
2.5\times 10^{-3}$\msunyr is $\sim 500$ times larger than the standard
Shu's value of $4.3\times 10^{-6}$\msunyr for 20~K gas. It is,
however, close to the characteristic free-fall rate for the clump as a
whole, ${\dot M}_{\rm c, ff}= 2.9 \times 10^{-3}$\msunyr (also plotted
on Fig.~1 for comparison). The large characteristic accretion rate
predisposes the clump to massive star formation, although this
tendency can be weakened or even suppressed by various factors,
including turbulence.

Turbulence reduces the total rate of mass accretion onto all stars as
long as it provides a significant global support to the clump. It is
therefore not surprising to find a smaller stellar mass accretion rate
in the case with turbulence (Model HD) compared to the base model
(BASE), especially at early times (see Fig.~1). The rate increases as
the turbulence decays, reaching values of order $2 \times
10^{-3}$\msunyr, not far below the characteristic free-fall rate
${\dot M}_{\rm c, ff}$, at $t\sim 0.5$~Myr, which is comparable to
both the free-fall time at the average density of the clump and the
crossing time for the initial turbulence.  Thereafter, the clump
dynamics is dominated by global collapse, as in the non-turbulent
model BASE. The global collapse is illustrated in the first panel of 
Fig.~2, which plots
the mass distribution and velocity field at the end of the simulation
($t=\ 3~t_{\rm ff,0}$) on a slice through the most massive star. As
expected, the dense materials are collected preferentially near the
bottom of the gravitational potential well, where they are fed by
the clump matter that falls towards the bottom. 

The total star formation rate is lowered further by magnetic fields
and outflows. Since the central plateau region of the clump has a 
mass-to-flux ratio that is significantly larger than the critical value,
the magnetic field there is not able to reduce the initial star formation
rate significantly. However, the magnetic field becomes more important 
at late times, after the initial turbulence has decayed substantially and
the more strongly magnetized clump envelope begins to collapse. The
ensuing global collapse is retarded by the magnetic field in the
cross-field direction. The retardation reduces the total stellar
accretion rate by a moderate factor of~2-3.

The global collapse is even less prominent when outflows are included
(the second panel of Fig.~2). The outflows have apparently supplied 
enough energy (and
momentum) into the clump that the bulk of its material is prevented
from either collapsing towards a global center or settling along field
lines into a thin sheet (although some degree of flattening in the
mass distribution is still evident). The feedback has kept the total
mass accretion rate at only $10\%$ of the characteristic free-fall
rate (shown in Fig.~1 for comparison). The global support directly 
affects the properties of the massive stars that are formed in the 
clump, as we show next.

\section{Massive Star Formation}
\label{massivestars}

In this section, we will concentrate on massive stars, defined
somewhat arbitrarily as those sink particles with masses more 
than 10~\msun. Our concentration on these stars is motivated by 
the fact that their formation mechanism is still under debate. We
postpone a discussion of the properties of the lower mass stars 
to another publication. 

\subsection{Properties of Massive Stars}

Massive stars are formed in all of the four simulations listed in
Table~1. In the limit of no turbulence (Model BASE), all of the
accreted mass goes to a single object at the center.  At the end of
the simulation ($t=2$), the object has grown to more than 400~\msun,
which is clearly unphysical. The addition of an initial turbulence in
Model HD enabled the formation of eight massive sink particles by
$t=3$. The time evolution of the positions of these particles are
shown in Fig.~3. There are two features worth noting. First, only five
lines are clearly distinguishable in the plot. This is because the
first 4 of the 8 stars are formed close together both in space and in
time; they stay close together at all times, so that their
trajectories are indistinguishable.  For our main purpose of
evaluating mass accretion rate, we will treat them as a single
object. Second, the remaining four massive stars are formed at
different times and locations.  Two of them join the first object at
later times to form a tight group, while the other two merge into
another system. We refer to the merged systems as group A and B
respectively, and the massive stars formed at different times in each
group as different generations.  For example, A1 is the first
generation of massive stars to form in group A. We will refer to all
massive stars in each generation as ``a massive object'' (because they
form at essentially the same time and same location, and partition of
mass between the stars depends somewhat on the sink particle
treatment).
Some properties of the massive objects are listed in Tables~3-5 
at three different times $t=3$, $4$ and $5.525$. 

There are drastic differences between the massive objects formed in
different simulations. At a relatively early time $t=3$ ($\sim
0.6$~Myr, about one global free fall time; Table~2), 5 massive objects
(8 massive stars) form in the HD Model, with a total mass of 131~\msun,
which is 37\% of the stellar mass. The presence of a magnetic field in
Model MHD reduces the number of massive objects (stars) to 3 (4), and
their total mass to 73.6~\msun, although the mass fraction of the
massive stars remains similar (45\%). When outflows are turned on
(Model WIND), the formation of massive stars is suppressed up to
this time, with no stars more massive than $10$~\msun. 

Massive stars do form at later times in Model WIND. At $t=4$, there
are two massive stars (16.9 and 10.5~\msun each) formed, with a
total mass of 27.4~\msun, which is much less than the total mass of
massive stars formed at the same time in Model MHD (183~\msun).  The
difference re-enforces the notion that the outflows retard massive
star formation. The massive stars do continue to grow with time,
however. They attain $46.4$ and $16.7$~\msun respectively near the end 
of the simulation at $t=5.5$ (about two global free fall times). They 
are each joined by another massive star, forming two loose groups.
Together, the four massive stars contain 86.9~\msun, accounting for
35\% of all the stellar mass in the cluster.

\subsection{Formation of the Most Massive Object} 

How do massive stars form in our simulations? To address this
question, we will concentrate on the most massive object formed 
in each simulation, starting with the simpler, HD Model that has
turbulence but neither magnetic fields nor outflow feedback.

\subsubsection{Model HD} 
%
%

The most massive object, HD-A1 in Table~2, formally reached a mass of
61.1~\msun at the end of the simulation ($t=3$ or 0.63 Myr). It is
among the first objects to form, at a time $t \sim 0.2$~Myr. The
average mass accretion rate is therefore ${\dot M}_{\rm ave}\approx
1.5\times 10^{-4}$~\msunyr. The instantaneous accretion rate for this
object is shown in Fig.~\ref{massivestar}.  It quickly rises to a
plateau of $\sim 10^{-4}$~\msunyr, before jumping at $t\sim
0.5$~Myr to a second plateau that is $\sim 2.5$ higher. About equal
amounts of mass ($\sim 30$~\msun) are accreted in the two plateau
phases. During the first plateau phase, the stellar mass increased by
an order of magnitude, while the mass accretion rate stayed more or
less constant. This behavior is characteristic of core collapse (such
as the collapse of the singular isothermal sphere where the mass
accretion rate is a constant, Shu 1977) rather than the Bondi-Hoyle
type accretion which depends sensitively on the stellar mass.  The
rapid increase in mass accretion rate around $\sim 0.5$~Myr is not
triggered by any sudden increase in stellar mass, which again points
to an external control of the mass accretion rate; it is caused by the
onset of global collapse of the clump.

To examine the nature of the accreting flow before the global collapse
in more detail, we plot in Fig.~\ref{HDYslice} the mass distribution,
velocity field and gravitation potential at $t=1.5$, when the eventual
massive object A1 is still in an early stage of accretion. It is
embedded in a dense filament, which is produced by converging flows in
the turbulence (see also Banerjee, Pudritz \& Anderson 2006). Here,
the inclusion of self-gravity in the calculation is crucial. It
allows the converged material to stay together, rather than
re-bouncing and dispersing quickly; in other words, the self-gravity
softens the high-speed collision that creates the filament.

The filament feeds the object at a high rate of $\sim 10^{-4}$
\msunyr, some 20 times the standard Shu's rate for an singular 
isothermal sphere (SIS). As stressed by Banerjee et al. (2006), 
the high accretion rate is due to the rapid formation of 
non-equilibrium structures (dense filaments) that are fed 
continuously by converging flows (and are thus very different 
from the equilibrium SIS). Their existence does not depend on 
the stellar objects embedded in them; rather, the objects are 
produced by the runaway collapse of the densest parts of the 
filaments, often at the intersections with other filaments. 
In this sense, the formation of the most massive object in this 
early phase is not due to competitive accretion, since the 
gravity of the object is not the primary driver of the mass 
accumulation in the filaments; there are few objects to compete 
against at this stage of star formation in any case. It may 
be viewed as a special version of core collapse, in which the 
``core'' is an over-dense, 
highly out-of-equilibrium filament that existed before the object and
that continues to grow out of the converging flow even as part of it
disappears into the collapsed object. The removal of dense gas from
the filament via star formation also promotes further mass
accumulation in the filament, by reducing its internal pressure and
thus the resistance to further compression. In this picture, the
massive star formation is simply part of the rapid ``core'' formation
process. The use of sink particles in our simulation enables us to go
beyond the calculations of Banerjee et al. (which stopped before
significant mass accretion onto the stellar object) and follow the
mass accretion for a long time, and to reveal a second phase of even
more rapid accretion, driven by global collapse.

The global collapse at the end of the pure hydro simulation (t=3) was
already shown in Fig.~2. It is further illustrated in 
Fig.~\ref{HDYslice}, which
shows the distribution of mass, velocity field and contours of
gravitation potential in a sub-region centered on the most massive
object, with the positions of stars superposed. It is clear that the
global collapse has supplied the central region near the bottom of the
gravitational potential well with plenty of dense material, which
fuels both the enormous rate of total stellar mass accretion (close to
the characteristic free-fall rate, see Fig.~1) and the high accretion
rate of the most massive object, which is but one of many stars in the
region.  In this crowded region, competitive rather than core
accretion is likely at work. The reason is that the stars in this
region are not enveloped by permanent host cores, because the dense
gas in the region is constantly drained onto stars and constantly
replenished by the global collapse. The most massive object is 
located near the minimum of the global potential well and accretes 
the global collapse-fed material at a rate higher than any other 
object. Nevertheless, its accretion rate is only $\sim 10\%$ of the 
total rate. In other words, the vast majority of the globally collapsing 
flow is diverted to stars other than the most massive object. Its 
accretion rate is large ($\sim 2.5\times 10^{-4}$~\msunyr) during 
this late phase only because of an even larger global collapse rate.

A common theme of the early and late phases of rapid accretion is that 
the dense gas that feeds the most massive object at a high rate is 
gathered by an agent external to the object. It is the converging 
flow set up by the
initial turbulence in the early phase and the globally collapsing flow
in the late phase. In this aspect, the formation mechanism is closer
to core collapse than to competitive accretion. One may plausibly
identify the dense filament in the early phase and the dense region at
the bottom of the global gravitational potential well in the late
phase as a McKee-Tan core (McKee \& Tan 2003). However, the ``cores'' 
so identified are
transient objects that are not in equilibrium. They are evolving
constantly, with mass growing (from converging or collapsing flow) and
depleting (into one or more collapsed objects) at the same time.  The
replenishment of dense gas in the ``core'' is the key to the long
duration of high mass accretion rate for the most massive object,
which is many times the free fall time of the small dense ``core'' 
and comparable to the global collapse time of the clump as a whole. 
In other words, there is no hint of any drop in the mass accretion 
rate after a finite reservoir of core material gets depleted over 
a (short) core
free-fall time. Indeed, the nearly constant or increasing accretion
rate for the most massive object highlights the important issue of 
when and how the mass accretion is terminated; there is an urgent 
need to tackle this issue if the final stellar mass is to be 
determined. We will return to a more detailed discussion of the 
nature of the massive star's high accretion rate in the absence 
of magnetic fields and outflow feedback in \S~\ref{unregulated}. 

\subsubsection{Model MHD}
\label{MHD}

The effect of the magnetic field in Model MHD on the mass accretion
rate onto the most massive object is relatively modest (see
Fig.~\ref{massivestar}). As in the HD case, there are two plateau
phases. The first starts somewhat later than that in the HD case,
indicating that the initiation of star formation in general and
massive star formation in particular is somewhat delayed, as expected,
because of magnetic cushion of turbulent compression. Once started,
the accretion rate onto the most massive object is in fact slightly
higher in the MHD case than in the HD case, indicating that the
magnetic field does not significantly impede the initial phase of
rapid mass accretion, which is enabled by the dense structures formed
by turbulent compression. Indeed, dense structure formation is aided
by a strong magnetic field, which forces the turbulent flows to 
collide along the field lines.
When viewed in 3D, the dense structure resembles a warped, 
fragmented disk, with transient spirals that come and go. 
The spirals indicate significant levels of rotation on 
relatively small scales, which may hinder the stellar mass 
accretion due to centrifugal barriers. 

The centrifugal barrier may be weakened or even removed by magnetic
braking. Evidence for the braking is shown in Fig.~\ref{MHDYslice},
which shows a magnetic bubble driven from the central region, where
active mass accretion is occurring. Magnetic braking driven bubbles
provide a form of energy feedback that should accompany any star
formed in a magnetized cloud (e.g., Draine 1980; Tomisaka 1998; Mellon
\& Li 2008; Hennebelle \& Fromang 2008) but is absent from purely
hydrodynamic models. We expect this form of feedback to become more
important with a higher spatial resolution, since the rotating
material closer to a star would be able to wrap the field lines more
quickly (until the flux freezing under the ideal MHD assumption breaks
down). Indeed, the protostellar outflow feedback may be viewed as a
special form of this magnetic bubble-driven feedback, since the
outflow is thought to be driven by the field lines that are forced to
rotate rapidly by the circumstellar disks (K\"onigl \& Pudritz 2000;
Shu et al. 2000).

Despite the feedback through the magnetic bubbles, large-scale
collapse does occur at late times.  When enough mass has been
accumulated near the bottom of the gravitational potential well, the
self-gravity of the accumulated gas overwhelms the magnetic tension
forces, leading to a rapid cross-field collapse and an elevated,
second phase of rapid mass accretion. The mass accretion is
facilitated by the magnetic braking, which enables the high angular
momentum material to sink closer to the center than it would be in the
absence of the braking. The collapse is not as global as in the HD
case, however, as shown in Fig.~\ref{infall}. At the end of the MHD
simulation ($t=4$), the bulk of the clump material infalls towards the
bottom of the potential well at a speed of order twice the sound speed
or less, unlike in the HD case, where the infall is faster and more
widespread. The collapse remains significantly retarded by the
magnetic tension except deep in the potential well.
%
%

\subsubsection{Model WIND}
%
%

The combined effect of outflows and magnetic fields on the formation
of the most massive object is much greater than that of magnetic
fields alone. From Fig.~\ref{massivestar}, we find that, even though a
massive star of more than 45~\msun was eventually formed at the end of
the WIND simulation, it took nearly $10^6$~yrs to accrete the mass,
with an average rate of only $\sim 5\times 10^{-5}$~\msunyr.  The
accretion rate at early times is even lower, with a value of $\sim
2\times 10^{-5}$~\msun during the initial third of the time; it
increases to $\sim 4\times 10^{-5}$~\msunyr during the second
third. These rates are about a factor of 5 lower than those in the HD
and MHD models at comparable times. The question is: why does the most
massive object accrete at such a low rate?

The basic reason is of course the outflows in the WIND model, which
are much more powerful than the magnetic braking-driven bubbles that
exist in the MHD case. The outflows modify both the mass distribution
and velocity field, and thus the outcome of the gravitational
dynamics. As a result, the most massive object forms at a somewhat
different time and location. Despite the additional physics in both
MHD and WIND models it is initially still embedded in a dense
filament, as it is in the HD case. As discussed earlier, the filament
is the key to the first, turbulence compression-induced phase of rapid
mass accretion in the HD and MHD models. In the WIND model, accretion
from the filament is reduced in two ways. First, there are several
stars formed in the filament. They all emit outflows which tend to
break the filament into smaller segments. Second, the outflows emerge
more easily perpendicular to (rather than along) the filament. Once
they break out, they blow against the converging flow and slow down
the mass accumulation that feeds the growth of the filament in the 
first place.

As the outflows propagate to large distances, a fraction of their
momentum is lost through the computation boundary. The remaining
fraction is deposited in the lower density region that surrounds the
denser region of active star formation near the bottom of the
potential well. The momentum deposition allows the bulk of the clump
material to be supported against rapid global collapse. The absence of
a global collapse is illustrated in Fig.~\ref{WINDYslice}, which shows
a rather chaotic velocity field that involves both infall and
outflow. It is further quantified in Fig.~\ref{infall}, which shows
that the
mass-weighted infall speed toward the most massive object is subsonic
over most of the volume, except within a radius of about 0.2~pc, where
the speed becomes two to three times the sound speed (still much lower
than the local free-fall speed). The slower infall leads to a smaller
amount of dense gas accumulating near the bottom of the potential
well. The accumulated dense gas is also more fragmented because of
interaction with outflows.  Both factors limit the mass accretion rate
onto the most massive object, especially at late times.

As in the HD and MHD cases, the most massive object is not formed out
of a pre-existing dense core. The case against the pre-existing dense
core picture is stronger in the WIND case, because it takes longer
(nearly two global free-fall times) to accumulate the final mass,
whereas a pre-existing dense core should collapse and exhaust its mass
in a local (core) free-fall time, which should be a very small
fraction of the global free-fall time. The slow accretion rate and
long accretion time are clear evidence that the formation of the most
massive object is controlled by the global clump dynamics in our WIND
simulation.

\section{Discussion}
\label{discussion}

%
%
%
%

\subsection{Unregulated Clump-Fed Massive Star Formation}
\label{unregulated}

Massive stars form quickly in our simulated parsec-scale clump in the absence 
of any magnetic field or outflow feedback (see Model HD). They are 
fueled by high mass accretion rates from either the dense filaments that 
are formed by turbulent compression or the clump-wide collapse due to 
the global turbulence dissipation. In both cases, the dense material that is 
depleted onto the massive star is constantly replenished. The replenishment 
is illustrated in the left panel of Fig.~\ref{starcore}, where the mass of 
the most massive object is plotted as a function of time, along with the 
mass of the gas within a ``core" of 0.1~pc in diameter around the 
object; the size was chosen to coincide with the fiducial value that 
McKee \& Tan (2003) used to define their 
``turbulent cores." It is clear that the 
``core" so defined has an initial mass that 
is well below the final mass of the most massive object, and thus cannot 
supply all of the mass of the object. As the object gains mass, the
mass of the ``core" stays roughly constant or even increases (rather 
than decreases), which 
demands that the core mass be replenished or fed from the surrounding 
clump. We are thus motivated to term this model the ``clump-fed massive 
star formation" model. 

Our ``clump-fed massive star formation'' model (CF model for short) 
contains elements of the two widely discussed 
scenarios for massive star formation in the literature: the
turbulent core model of McKee \& Tan (2003) and competitive accretion
model of Bonnell et al. (2003). It has in common with the turbulent 
core model in that the mass accretion rate onto the massive star is
not primarily determined by the star itself, but rather by the
properties of the pre-existing gas that produces the seed of the star in
the first place and that continues to feed the stellar growth; in
other words, if the massive star were to be removed prematurely,
another seed would be produced in its place and grow to a high
mass. It is the properties of the fueling gas that determine 
the stellar mass accretion rate and thus the stellar mass. 
McKee \& Tan envisioned this gas to be a pre-existing
massive turbulent core. It collapses to form the massive star, with a
mass accretion rate that depends on the core structure.  For example,
if the turbulent core has a density profile of $r^{-1.5}$, the mass
accretion rate would increase linearly with time $t$ (McKee \& Tan
2003). In our CF model, the pre-existing gas is the
cluster-forming clump, which produces the transient dense material
that feeds the massive star at a high rate through two mechanisms: 
(1) the collapse of the dense filaments produced by turbulent 
compression, as already emphasized by Banerjee et
al.~(2006), and (2)~global collapse, driven by the turbulence decay 
in our simulations, but can in principle be caused by an external 
compression 
as well. The first mechanism depends on the detailed properties of 
the initial turbulence in the clump, which are not well constrained 
observationally. The second mechanism depends on the dissipation 
of (supersonic) turbulence that is inevitable, and should be more 
robust. 

Our unregulated CF model has in common with the competitive 
accretion model in that a massive star can form even in the absence 
of a pre-existing turbulence-supported massive core and
that the global gravitational potential and dynamics of the clump 
play an important, even the dominant, role in massive star formation. 
The latter is especially true at late times, when the global collapse 
feeds mass at a high rate to the compact region near the bottom of 
the potential well, where a large number of stars (including massive 
stars) are already present. In the absence of stellar feedback, 
competition for this clump-fed 
material in the crowded region is unavoidable. The most massive object
tends to grow the fastest, mainly because it typically locates closest
to the center of the potential, as emphasized by Bonnell et
al. (2007). 
Although the numerical results of our grid-based AMR hydro 
simulations are in broad agreement with those of SPH simulations of 
Bonnell et al. (2003), we differ from their
interpretation of the results regarding the formed stars, particularly
as it relates to massive star formation. When a massive star is
formed, the vast majority of the clump mass remains in the gas. 
The high accretion rates of the massive stars at late times of our
hydro simulation derive directly from the global gas collapse. Even if
there were no stars near the center of the collapse (or more likely
the mass accretions onto individual stars are terminated by the
stellar feedback), (other) massive stars can still form from scratch,
as long as the collapse delivers mass to the center at a high enough
rate. So we argue it is predominantly the structure and dynamics
of the gaseous component that set the relevant physics in forming 
the massive stars rather than the properties of the stars made
previously. 

A moderately strong magnetic field (corresponding to the observationally
inferred dimensionless mass-to-flux ratio of a few) does not qualitatively 
change the clump-fed picture for massive star formation by itself. As 
discussed in  
\S~\ref{MHD}, the magnetic field has relatively little effect on the 
high mass accretion rate onto the most massive object induced by the
initial turbulent compression and filament formation. Indeed, it is conducive
to filament formation by guiding the converging flows to collide along the 
field lines. The magnetic field does slow down 
the global collapse-induced rapid accretion by a modest factor of a few. 
The basic ingredients of the CF model remain, however,
as can be seen from the middle panel of Fig.~\ref{starcore}. Specifically, 
the initial core mass is still much less the final mass of the most
massive object, and most of the mass that goes into the object still 
needs to be replenished or fed from the surrounding clump. We conclude 
that the feeding processes for massive star formation are only weakly 
regulated by the magnetic field, perhaps because the field is only 
moderately strong (i.e., the clump is moderately supercritical), and it 
resists the feeding passively rather than actively (unlike the
outflows, see below). The magnetic field
does change the dynamical coupling between different parts of the 
clump, an important aspect of cluster and massive star formation 
that we plan to return to in a future investigation. 

\subsection{Outflow-Regulated Clump-Fed Massive Star Formation}

 Massive star formation through rapid accretion of the mass that is fed from 
 outside the small region surrounding the forming star  
 is particularly vulnerable to outflow feedback. This is because the
 feeding is directly opposed by the feedback. In the case of the  rapid 
 accretion fed by turbulent compression and filament formation, 
 the filament is quickly chopped up into small segments by the outflows 
 driven by the stars formed in it (see the outflow movie in the electronic 
 version of the paper). The outflows also slow down any further mass 
 accumulation in the filament after star formation is initiated. As a result, 
 this mode of turbulent compression-fed massive star formation is 
 strongly regulated, perhaps even choked off completely, by outflow 
 feedback. 
 
 In the case of the rapid accretion fed by global collapse, the infall 
 is countered by the outflows on all scales, especially on the global 
 clump scale. The additional global support provided by the outflow 
 feedback can reduce the total star formation rate by a large factor. 
 This reduction affects the formation of all cluster members, 
 especially the massive stars. This is because the massive stars
 receive a larger fraction of the collapse-fed material, and are thus
 more sensitive to the change of global clump dynamics. They 
 also tend to complete their formation toward the end of cluster 
 formation (even though their seeds tend to be among the first 
 objects to form), making them more prone to the accumulative 
 influence of multiple generations of outflows that precede their 
 eventual formation. Since the outflows are believed to be driven 
 by the release of gravitational binding energy from mass accretion 
 in one form or another (Konigl \& Pudritz 2000; Shu et al. 
 2000), and most of the accreted mass goes to low-mass (rather 
than high-mass) stars for a Salpeter-like IMF, a large 
 fraction if not the bulk of the outflow feedback should come from 
 the low-mass stars. In this sense, the formation of low-mass stars 
 in a dense clump can profoundly influence the formation of massive 
 stars in the same clump, through their feedback on the clump dynamics. 
 Whether massive stars eventually form or not in a dense clump depends 
 on the extent to which all the outflows in the clump collectively 
 regulate the global collapse and slow
 down the star formation. If the total star formation rate is 
 reduced, for example, below the fiducial minimum rate for massive
 star formation, $10^{-4}$\msunyr, no massive stars would 
 form at all. In the opposite extreme where the outflows are
 weak and the feedback is not strong enough to reverse the
 global collapse, stars (especially massive stars) would form
 quickly, as in our pure hydro simulation. 
 
 The degree of outflow regulation will depend on the properties of 
 the outflows, including their strengths and degrees of collimation, 
 both of which are somewhat uncertain. 
 What we have demonstrated explicitly through numerical simulation 
 is that for well-collimated jets of reasonable strength the outflow 
 feedback can prevent rapid global collapse, and keep the total star 
 formation rate an order of magnitude below the characteristic 
 free-fall rate. Massive stars do eventually form in our simulation 
 that includes outflow feedback. It demonstrates that outflows can 
 strongly regulate massive star formation, but do not necessarily 
 quench it completely, especially in dense massive clumps with a
 high characteristic free-fall rate ${\dot M}_{\rm c,ff}$. For
 such clumps, even when the bulk of the clump material is supported, 
 a small fraction can still percolate down the global gravitational 
 potential well, feeding the formation of massive stars near 
 the center at a high enough rate. 

 The fact that our outflow-regulated massive star formation remains 
 clump-fed rather than core-fed is illustrated in the right panel 
 of Fig.~\ref{starcore}. As in the HD and MHD models that do not  
 have outflow feedback, the mass of the 0.1~pc-sized ``core'' that 
 surrounds the most massive object in the WIND model is initially 
 smaller than the final mass of the object, and does not decrease 
 monotonically as the object accretes. Again, the bulk of the 
 accreted stellar material must come from outside the ``core,'' 
 which is an essential feature of our new scenario of outflow-regulated 
 clump-fed (ORCF) massive star formation. In this picture, the 
 parsec-scale cluster-forming clump, rather than a 0.1~pc-sized 
 turbulent core, is the basic unit for massive star formation. 
 As pointed out by Bonnell et al. (2007), a potential drawback of 
 the turbulent core model is that massive, turbulence-supported, 
 cores tend to fragment into many stars rather than collapse 
 monolithically, although radiative feedback from the rapidly 
 accreting massive stars can reduce the level of fragmentation  
 (Krumholz \& McKee 2008). Fragmentation is expected to be less 
 of a problem in our clump-fed picture of massive star formation, 
 since the material near the forming massive stars does not have 
 to be supported by a strong turbulence; it is typically in a 
 state of rapid collapse that feeds the growing massive stars 
 at a high rate, even though the clump as a whole may remain 
 supported by (possibly outflow-driven) turbulence and, perhaps 
 to a lesser extent, magnetic fields. If the parsec-scale dense 
 clumps are indeed the basic units of massive star formation,  
 how they form in the first place becomes an important problem 
 that deserves close theoretical and observational attention. 
%
%
 %
  
 \subsection{A Condition for Massive Star Formation in 
 Galaxy Formation Simulations}

Massive stars play a dominant role in galaxy formation and evolution. 
Our ORCF scenario suggests a rough criterion for their formation 
that can be used in global galaxy formation simulations that reach 
the scale of the cluster forming clumps but do not resolve their 
internal structure.  

The criterion is based on a threshold for mass accretion rate. A 
high mass accretion rate is needed not only to overcome the 
radiation pressure (Wolfire and Cassinelli 1987) and quench 
the development of HII region (Walmsley
1995), but also to satisfy observational constraints on the time scale
of massive star formation. Wood \& Churchwell estimated an age of
$\sim 10^5$~years for UC HII regions, which probably represent a
relatively late stage of massive star formation, when the bulk of mass
accretion has completed and the mass accretion rate becomes too low to
trap the HII region (Churchwell 2002). The majority of the stellar
mass may be accreted during the hot core (and perhaps hypercompact
HII) phase, which lasts for a time of order $10^5$~years or less
(Kurtz et al. 2000). The relatively short time scale is also
consistent with the dynamical times estimated for massive molecular
outflows (which are presumably driven by rapid mass accretion during
the main accretion phase, as in the case of low mass stars; e.g.,
Bontemps et~al.~1996), which are typically of order $10^5$ years or
less (e.g., Zhang et~al.~2005). If the time scale for massive star 
formation is indeed $\sim 10^5$~yrs or less, to form a star of
10\msun, a stellar mass accretion rate of ${\dot M}_{\rm cr} \sim
10^{-4}$~\msunyr or more would be needed. A minimum requirement
for massive star formation in a clump is that the characteristic
free-fall rate ${\dot M}_{\rm c,ff}$ of the clump be greater than 
${\dot M}_{\rm   cr}$.

The actual requirement will be more stringent. The accretion rate
onto massive stars ${\dot M}_{m*}$ is related to the
characteristic free-fall rate ${\dot M}_{\rm c,ff}$ of the dense clump
by two factors: ${\dot M}_{m*} = {\dot M}_{\rm c,ff}\ f_*\ f_{m*}$, 
where $f_*$ is the actual rate of star formation normalized
to the characteristic rate ${\dot M}_{\rm c, ff}$, and $f_{m*}$ the
fraction of the stellar mass accretion onto massive stars. We 
estimate the characteristic free-fall rate using the clump mass 
divided by the free-fall time at the average density
\begin{equation}
{\dot M}_{\rm c,ff} \sim (8 G/\pi^2)^{1/2} (M/R)^{3/2} \sim 1.9 \times
10^{-3} (M_3/R_{pc})^{3/2} M_\odot {\rm yr}^{-1}
\end{equation}
(where $M_3$ is the clump mass in units of $10^3 M_\odot$ and $R_{pc}$
the clump radius in units of parsec), which yields $2.5\times 10^{-3}
M_\odot$/yr for a $1200$~\msun clump of 1~pc in radius, consistent
with the value obtained in our reference model of non-turbulent
collapse.  The most uncertain factor in the above equation is perhaps
$f_*$, the star formation rate compared to the characteristic
free-fall rate.  It depends on the extent to which the clump is
supported globally.  Decaying turbulence can provide significant
global support within a turbulence decay time, when $f_*$ can be
significantly below unity.  Unless the star formation is terminated in one
decay time, $f_*$ will eventually increase to a value not far from
unity, as demonstrated in our HD model. A moderate magnetic
field does not change the global support fundamentally, reducing $f_*$
by a factor of only a few (see Model MHD). Together with the magnetic
field, protostellar outflows can reduce $f_*$ by an order of
magnitude, to values of order $\sim 0.1$. The value of $f_*$ may range
from about 0.1 (before the virial turbulence decays away or after it
is replenished) to close to unity (after the turbulence has decayed in
the absence of magnetic support and turbulence
replenishment). Krumholz \& Tan (2007) argued that the star formation
efficiency per free fall time is of order $10\%$ or less, implying
$f_* \leq 0.1$. Nakamura \& Li (2007) found a similarly low value for
the well-studied nearby clump of active cluster formation NGC
1333. Based on these results, we choose $f_* \sim 0.1$ as the fiducial
value. A smaller $f_*$ would make it more difficult to form massive
stars.

The value for the remaining factor $f_{m*}$ can be constrained both 
observationally and numerically.  If the Salpeter slope is universal
for the upper part of the IMF, then 
about 1/3 of the stellar mass must reside in stars more massive 
than 10~M$_\odot$ (assuming a lower cut-off at 
0.3~$M_\odot$). In this case, $f_{m*}\sim 1/3$, which is similar to
the ratios obtained in our simulations, according to Tables~2-4. To be 
conservative, we assume that all of the massive star accretion rate, 
${\dot M}_{\rm c,ff} f_* f_{m*}$, goes to a single star. If more than 
one massive star are fed at this rate, the requirement for massive 
star formation would be more stringent.

Taken together, the above considerations yield the following mass accretion 
rate for a massive star:
\begin{equation}
{\dot M}_{m*} \sim 6.4 \times 10^{-5} M_\odot/yr 
\left({f_*\over 0.1}\right) \left({f_{m*}\over 1/3}\right) 
\left({M_3\over R_{pc}}\right)^{3/2},
\end{equation}
which has to exceed $10^{-4} {\dot M}_{cr,-4} M_\odot/yr$ for massive star
formation to actually occur (${\dot
  M}_{cr,-4}$ is the critical mass accretion rate for 
massive star formation in units of $10^{-4}$\msun/yr). 
This yields a rough condition on the
ratio of the mass and radius of the clump for massive star formation: 
\begin{equation}
{M_3\over R_{pc}} \gtrsim1.4 \left({0.1\over f_*}\right)^{2/3} \left({1/3 
\over f_{m*}}\right)^{2/3} {\dot M}_{cr,-4}^{2/3}.
\label{condition} 
\end{equation} 

The significance of the above condition is that, if the fraction of 
the total stellar mass 
going into massive stars ($f_{m*}$) is constrained by observations, then 
whether a dense clump of a given mass and size (and 
thus given ${\dot M}_{\rm c,ff}$)  can form massive stars or not boils down to 
the total star formation rate, $f_* {\dot M}_{\rm c,ff}$.  It depends
on the clump 
dynamics, which is sensitive to the outflow feedback. The above condition 
is consistent with, for example, the sample
of massive star forming clumps associated with water masers of Plume
et al. (1997), which has a mean virial mass $M_3=3.8$ and mean radius
$R_{pc}=0.5$. Their ratio of 7.6 is comfortably above the fiducial
value of $1.4$ in equation~(\ref{condition}), indicating that massive
stars can form, even when the total rate of star formation is an
order of magnitude below the characteristic free-fall rate. A small 
value for the product $f_* f_{m*}$ ($\sim 1/30$) may be the reason 
for the massive star formation to occur predominantly in those 
special regions of molecular clouds---the massive dense clumps---that 
are both massive and compact.   

\subsection{Limitations and Future Directions}
\label{limitation}

The most severe limitation of the current work is perhaps the neglect
of radiative feedback. Radiative heating changes the clump 
fragmentation behavior, 
especially close to the forming stars (Krumholz et al. 2007; Bate 2009). This 
effect was mimicked to some extent by the relatively large sink particle 
merging distance adopted in our simulations (with a length of 5 cells
or $10^3$~AU), which 
suppresses fragmentation on the small (mostly disk) scale. Furthermore, 
if our ORCF scenario is correct, the formation of massive stars may be 
more sensitive to the global clump dynamics (which are less affected by 
the radiative heating) than the gas properties close to the 
stars. Nevertheless, we believe that the formation of massive stars 
will benefit from the suppression of fragmentation by radiative 
heating, especially near the bottom of the gravitational potential 
well of the clump, where the thermal Jeans mass is formally smaller 
than the typical stellar mass. On small scales, radiative pressure may 
slow down the mass accretion onto individual massive stars somewhat, 
although the accretion may be enhanced to some extent by non-ideal 
MHD effects, such as ambipolar diffusion, which are not included in 
our ideal MHD calculations.    

Another effect that we neglected was the HII region driven by the massive 
star's UV radiation. The expansion of HII regions provides a way to remove 
the clump gas, and perhaps terminate the cluster formation. It needs
to be included in future simulations that aim to model the entire 
history of cluster formation. 
Such studies may also need to include massive star winds, which are 
observed to have dramatic effects in some regions (e.g. the Carina
Nebula, see Smith \& Brooks 2008 for a review). They are the main 
alternative to the HII regions as the means for terminating the 
cluster formation. 

A further limitation is the periodic boundary condition used in our 
simulation. It precludes any communication between the 
cluster-forming clump and its 
surrounding environment. If there is energy injection into the dense clump
from the ambient medium, the externally supplied energy may aid the 
outflows in regulating the cluster and massive star formation. Large-scale
external compression may, on the other hand, lead to rapid clump 
collapse and massive star formation. In this case, the feeding of massive
stars may extend beyond the parsec-scale clump, and the massive star
formation may simply be part of the rapid clump formation. 

\section{Summary and Conclusion}
\label{summary}

We have carried out AMR-MHD simulations of massive star formation in
dense, turbulent, parsec-scale clumps of cluster star formation 
including sink particles and outflow feedback. We find that, without 
regulation by magnetic fields and outflows, massive stars form
quickly. They are fed at a high rate first by the converging flows 
in the initial turbulence and later by the global collapse induced 
by turbulence decay. A moderate magnetic field alone does not affect 
these feeding processes much. They are greatly modified, however, 
by a combination of protostellar outflows and magnetic fields. 
The outflows break up the turbulent compression-produced dense 
filaments that feed the massive stars at early times and stall 
the global collapse that fuel the massive star formation at 
later times. The outflow feedback is enhanced by a magnetic field, 
which links different parts of the clump together; the coupling 
makes the deposition of the outflow momenta in the clump more 
efficient. The magnetically-aided outflow feedback can in 
principle reduce the total rate of star formation below the critical 
mass accretion rate for massive star formation and suppress the 
massive star formation completely. In practice, whether massive 
stars form in a dense clump or not depends on the properties of 
the clump (particularly its mass and size) and the degree of 
magneto-outflow regulation of its star formation (see 
equation~[\ref{condition}]). For parsec-scale clumps of order 
$10^3$\msun, we have demonstrated explicitly through numerical 
simulations that the formation of massive stars is clump-fed 
and outflow-regulated. Additional simulations and analysis are 
needed to determine whether this new scenario of outflow-regulated 
clump-fed massive star formation is applicable to more massive 
and/or more compact dense clumps. In a companion paper, we will 
explore the effects of the outflow feedback on the lower mass 
cluster members. 

ZYL is supported in part by NASA (NNG05GJ49G) and NSF (AST-030768) 
grants, and FN by a Grant-in-Aid for Scientific Research of Japan 
(20540228).

\begin{deluxetable}{lllll}
\tablecolumns{5}
\tablecaption{Model Charateristics \label{tab:model}}
\tablewidth{\columnwidth}
\tablehead{
\colhead{Model}     & \colhead{Turbulence} & \colhead{Magnetic Field}
&\colhead{Outflow}   & \colhead{Stop Time} 
}
\startdata
BASE   & no & no  & no  & 2.0 \\
HD   & yes & no  & no   & 3.0 \\
MHD   & yes & yes  & no   & 4.0 \\
WIND  & yes & yes  & yes   & 5.525 \\

\enddata
\tablecomments{The stop time of each simulation is given in units of
the initial free-fall time at the clump center $t_{\rm ff}= 0.21$~Myr.}
\end{deluxetable}

\begin{deluxetable}{llllll}
\tablecolumns{6}
\tablecaption{Massive Objects at $t=3$ \label{tab:m3}}
\tablewidth{\columnwidth}
\tablehead{
\colhead{Name}     & \colhead{Mass} & \colhead{$t_{\rm form}$}
&\colhead{\#}   & \colhead{Location}  & \colhead{Notes} 
}
\startdata
HD-A1   & 61.1  & 0.96  & 4 & (0.66, 0.40, 0.48) & \\
HD-A2   & 22.6  & 1.36  & 1 & (0.66, 0.40, 0.48) & \\
HD-A3   & 10.1  & 2.03  & 1 & (0.68, 0.41. 0.48) & \\  
HD-B1   & 21.3  & 1.11  & 1 & (0.74, 0.48, 0.33) & \\ 
HD-B2   & 15.8  & 1.72  & 1 & (0.74, 0.48, 0.33) & \\ 
HD-ALL  & 131   &       & 8 &                    & 137 stars, 350\msun; 
37\% massive\\ \hline
MHD-A1  & 12.3 & 1.05  & 1   & (0.41, 0.34, 0.51) &  \\
MHD-A2  & 46.7 & 1.35  & 2   & (0.44, 0.38, 0.52) &  \\
MHD-B1  & 14.6 & 2.30  & 1   & (0.51, 0.65, 0.36) &  \\
MHD-ALL & 73.6 &       & 4   &                    & 86 stars, 
165~\msun; 45\% massive \\ \hline
WIND-ALL  & 0 &   & 0   &    & 72 stars, 69.7~\msun; no massive stars \\
\enddata
\tablecomments{The units for the stellar mass, formation time, and
  stellar location are the solar mass, the initial free-fall time 
  at the clump center, and the box size (2 pc). The 4th column 
  denotes the number of stars in each generation. The number of
  all stars more massive than 0.1~\msun, their total mass, and 
  the fraction of the total stellar mass in massive objects are
  noted in the last column.}
\end{deluxetable}

\begin{deluxetable}{llllll}
\tablecolumns{6}
\tablecaption{Massive Objects at $t=4$ \label{tab:m4}}
\tablewidth{\columnwidth}
\tablehead{
\colhead{Name}     & \colhead{Mass} & \colhead{$t_{\rm form}$}
&\colhead{\#}   & \colhead{Location}  & \colhead{Notes} 
}
\startdata
MHD-A1  & 15.0 & 1.05  & 1   & (0.32, 0.45, 0.51) &  \\
MHD-A2  & 110  & 1.35  & 2   & (0.44, 0.40, 0.51) &  \\
MHD-A3  & 12.5 & 3.13  & 1   & (0.46, 0.41, 0.51) &  \\
MHD-B1  & 45.8 & 2.30  & 1   & (0.46, 0.65, 0.39) &  \\
MHD-ALL & 183  &       & 5   &                    & 121 stars, 
337~\msun; 54\% massive \\ \hline
WIND-A1 & 16.9 & 1.16  & 1   & (0.38, 0.42, 0.50) & \\
WIND-B1 & 10.5 & 2.29  & 1   & (0.45, 0.68, 0.42) & \\
WIND-ALL & 27.4 &   & 2   &    & 123 stars, 134~\msun; 20\% massive\\
\enddata
\tablecomments{Units as in Table~\ref{tab:m3}.}
\end{deluxetable}

\begin{deluxetable}{llllll}
\tablecolumns{6}
\tablecaption{Massive Objects at $t=5.5$ \label{tab:m5}}
\tablewidth{\columnwidth}
\tablehead{
\colhead{Name}     & \colhead{Mass} & \colhead{$t_{\rm form}$}
&\colhead{\#}   & \colhead{Location}  & \colhead{Notes} 
}
\startdata
WIND-A1 & 46.4 & 1.16  & 1   & (0.47, 0.36, 0.51) & \\
WIND-A2 & 11.8 & 1.34  & 1   & (0.43, 0.33, 0.47) & \\
WIND-B1 & 16.7 & 2.29  & 1   & (0.49, 0.67, 0.44) & \\
WIND-B2 & 12.0 & 2.31  & 1   & (0.51, 0.67, 0.43) & \\
WIND-ALL & 86.9 &   & 4   &    & 183 stars, 250~\msun; 35\% massive\\
\enddata
\tablecomments{Units as in Table~\ref{tab:m3}.}
\end{deluxetable}

\begin{figure}
\epsscale{0.65}
\plotone{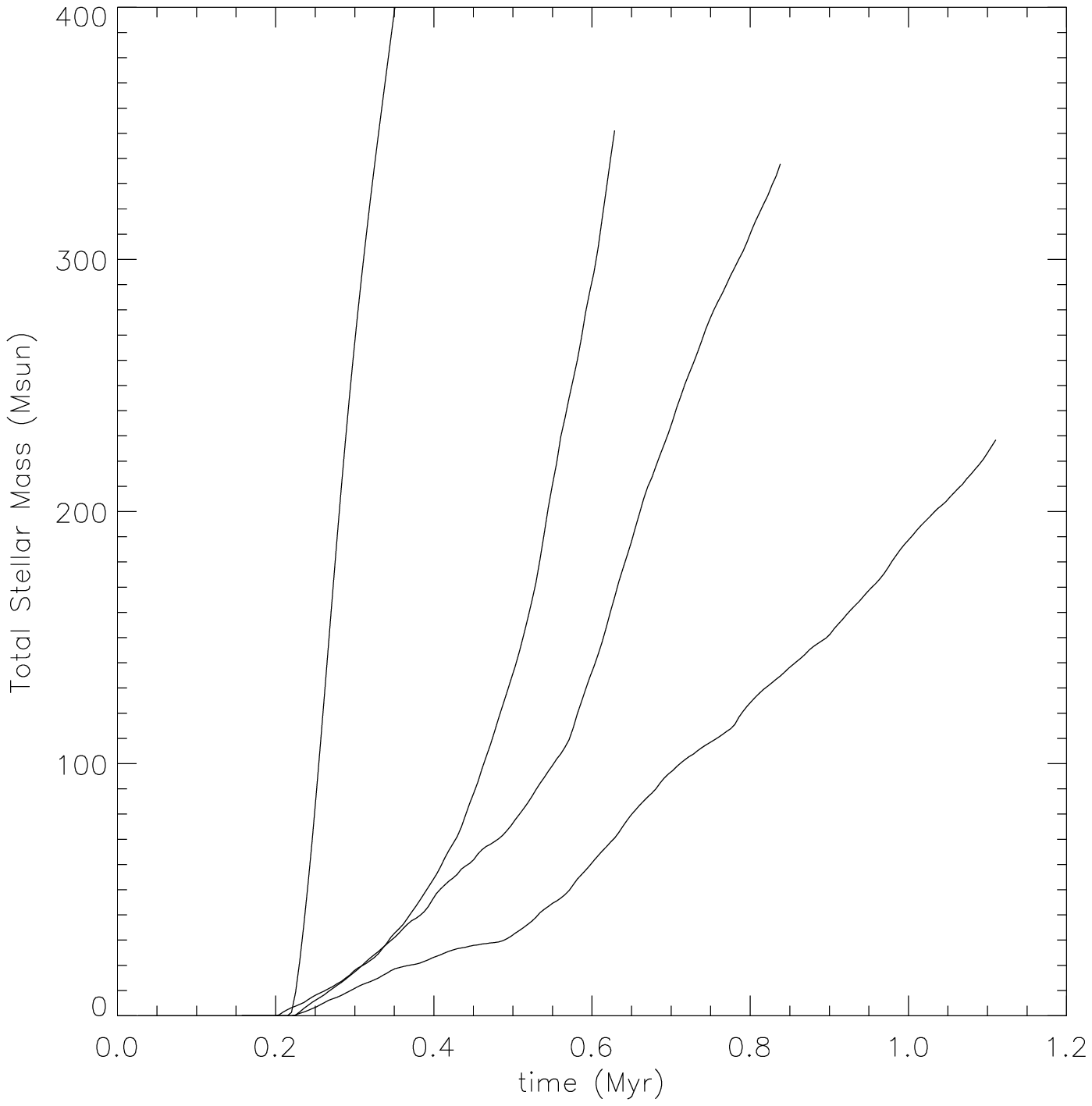}
\plotone{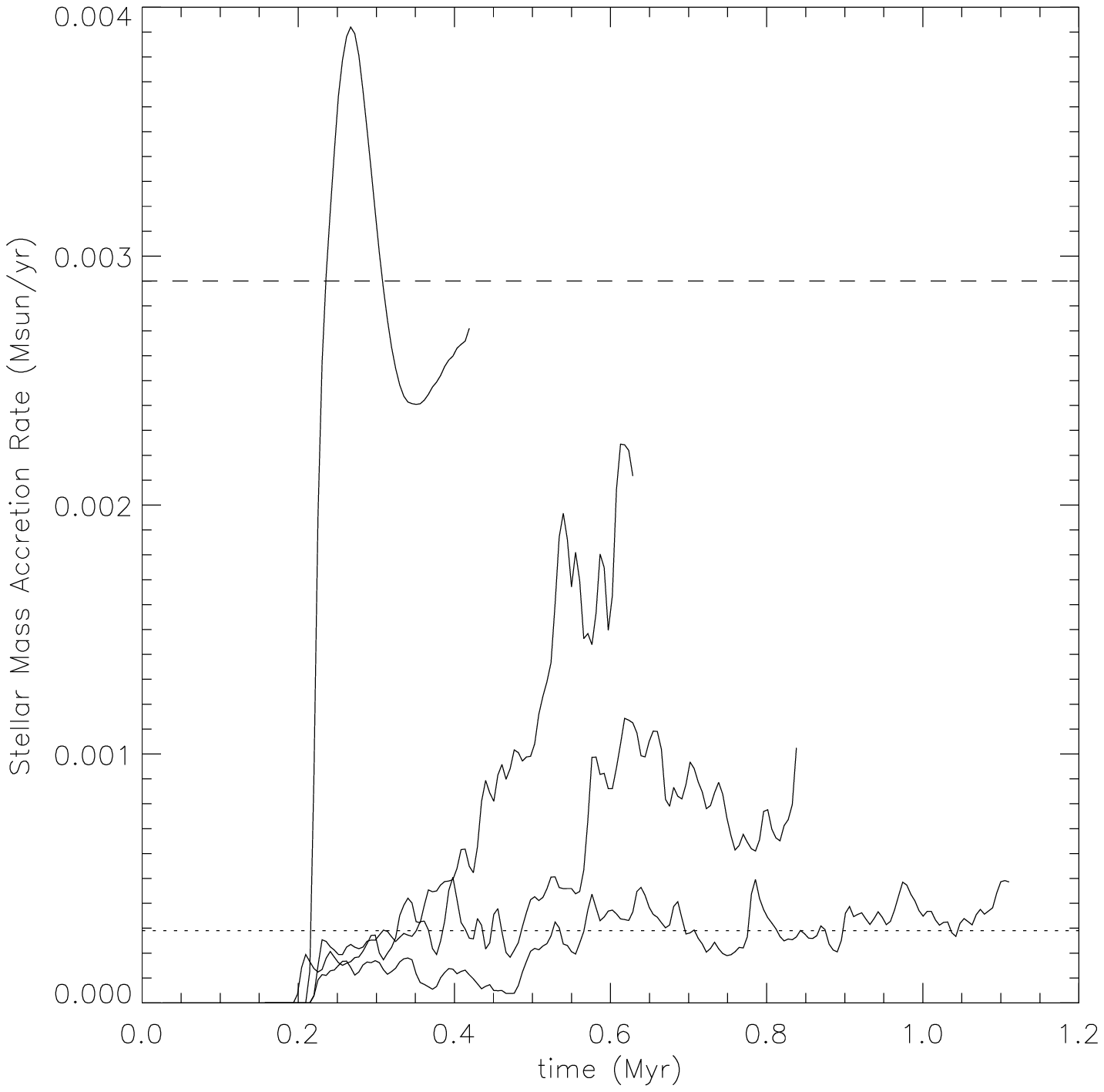}
\caption{Evolution of the total stellar mass (top panel) and stellar
 mass accretion rate (bottom panel) for all 4 models. In each panel, 
 the curves from upper-left to lower-right are for Model BASE, HD, 
MHD and WIND, respectively.}. 
\label{stellar}
\end{figure}

\begin{figure}
\epsscale{0.65}
\plotone{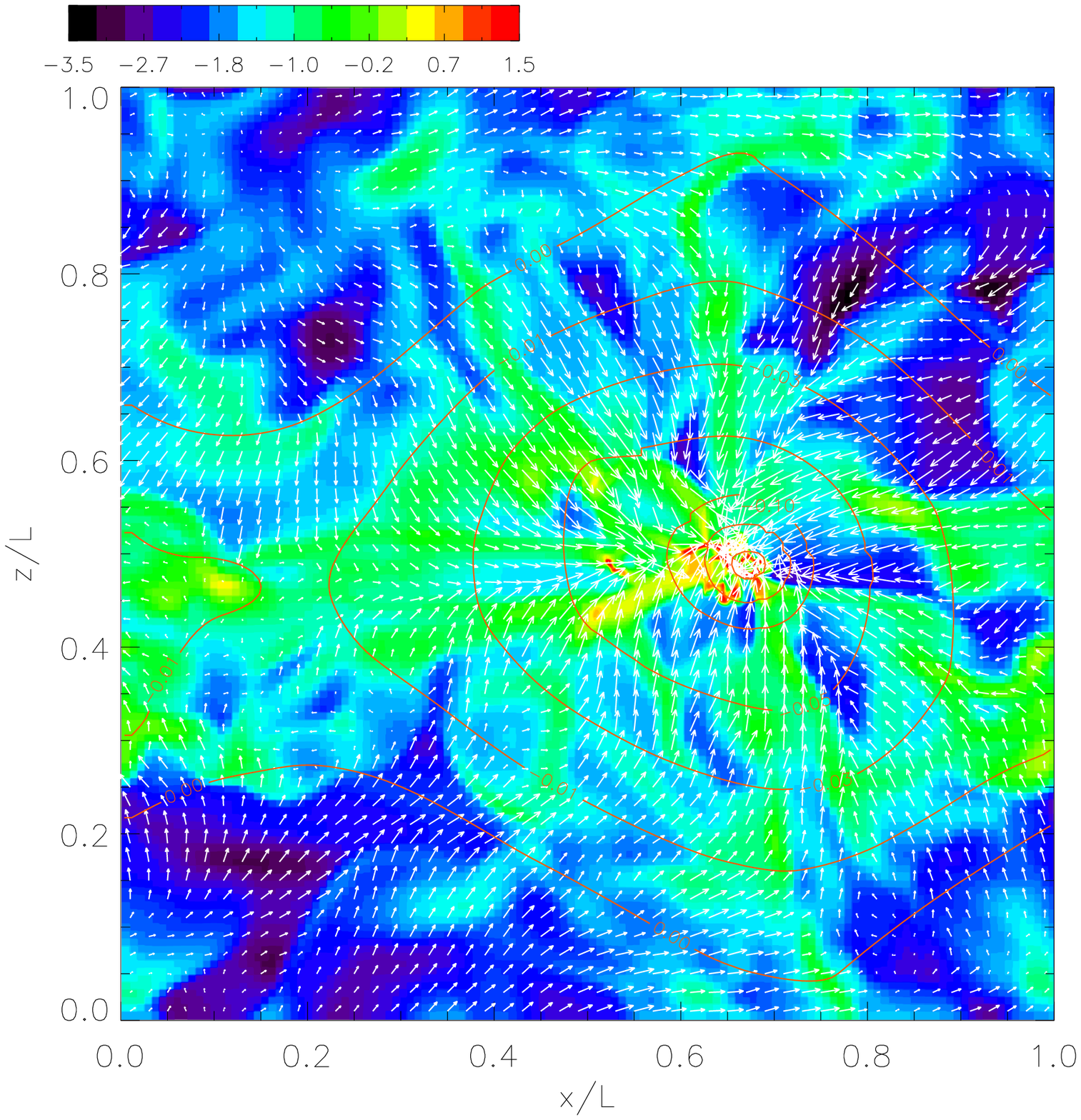}
\plotone{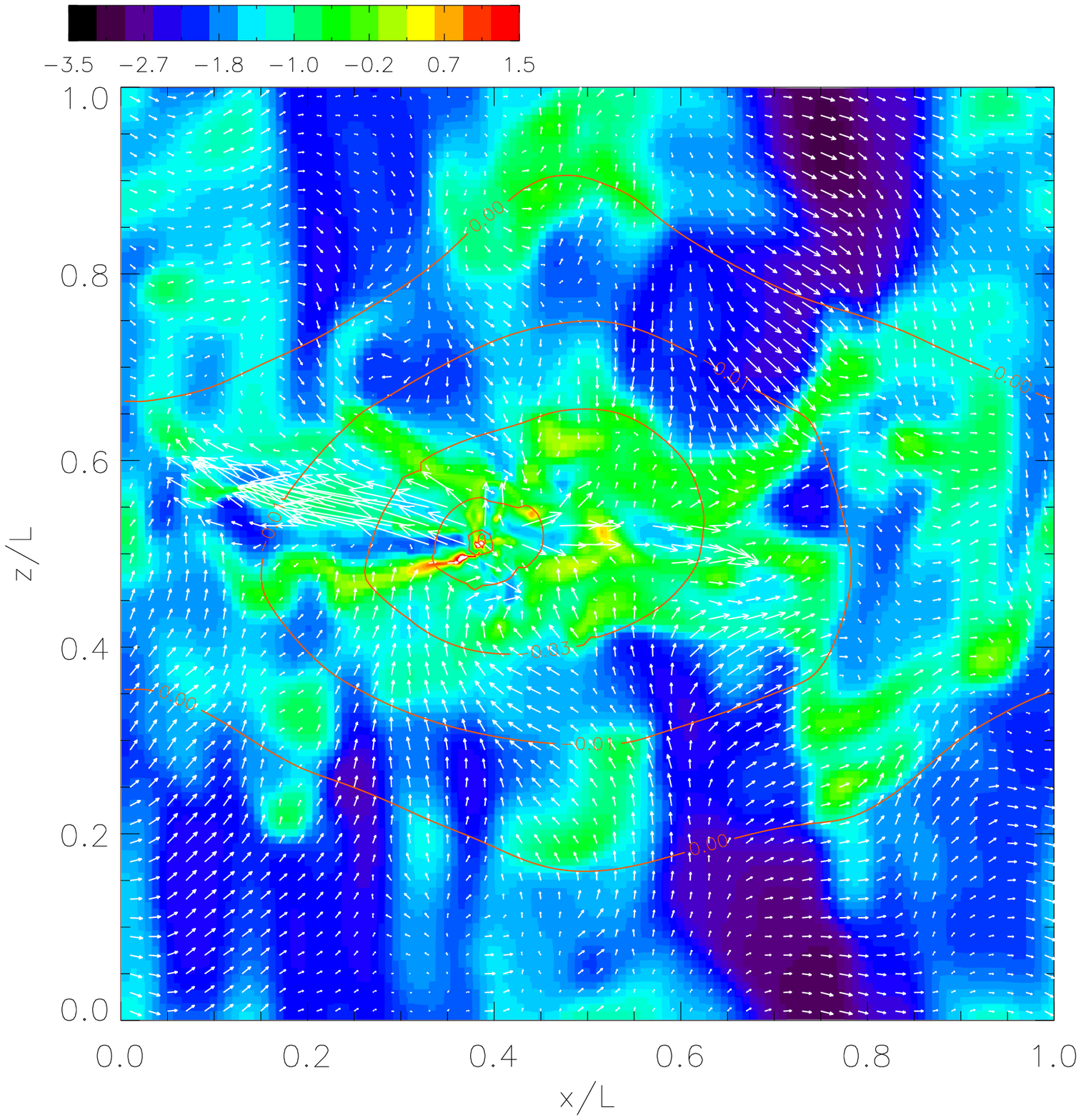}
\caption{Density slice through the most massive object for the HD and WIND 
model at t=3. Overplotted are the velocity arrows (white) and contours of 
the gravitational potential. The colorbar is for the logarithm of the 
density in units of the initial central density of the clump, and
length is scaled by the size of the simulation box $L$. } 
\label{SliceAt3}
\end{figure}

\begin{figure}
\epsscale{1.05}
\plotone{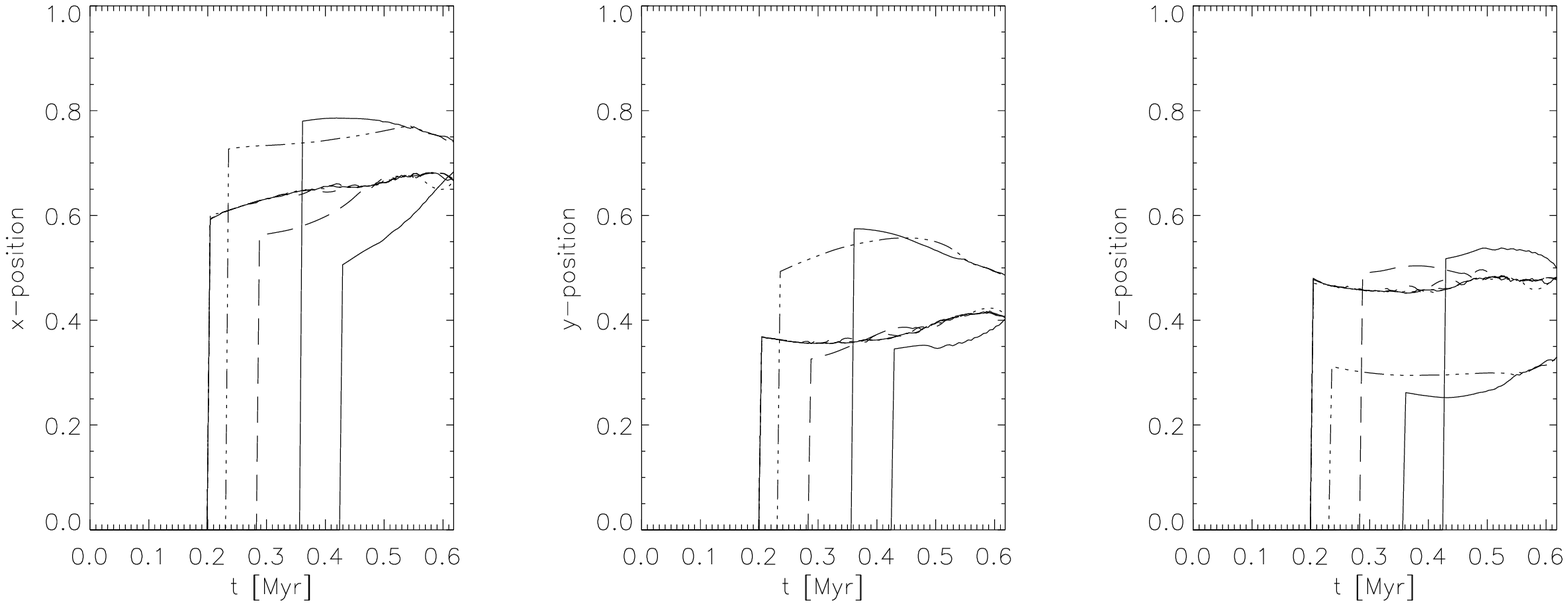}
\caption{Evolution of the positions of the massive stars in the HD Model,
 showing the increase in the degree of clustering at later times.} 
\label{position}
\end{figure}

\begin{figure}
\epsscale{0.65}
\plotone{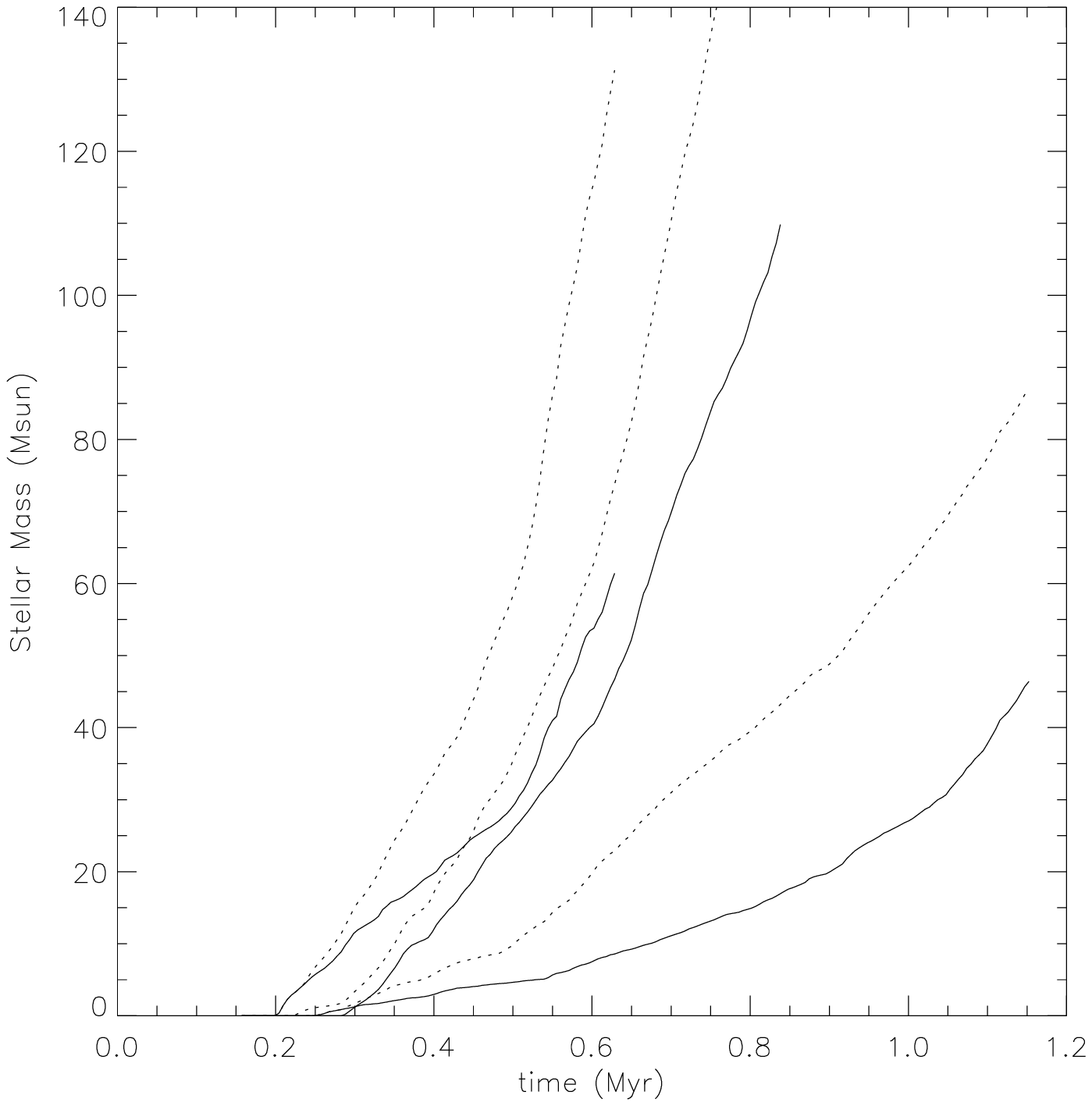}
\plotone{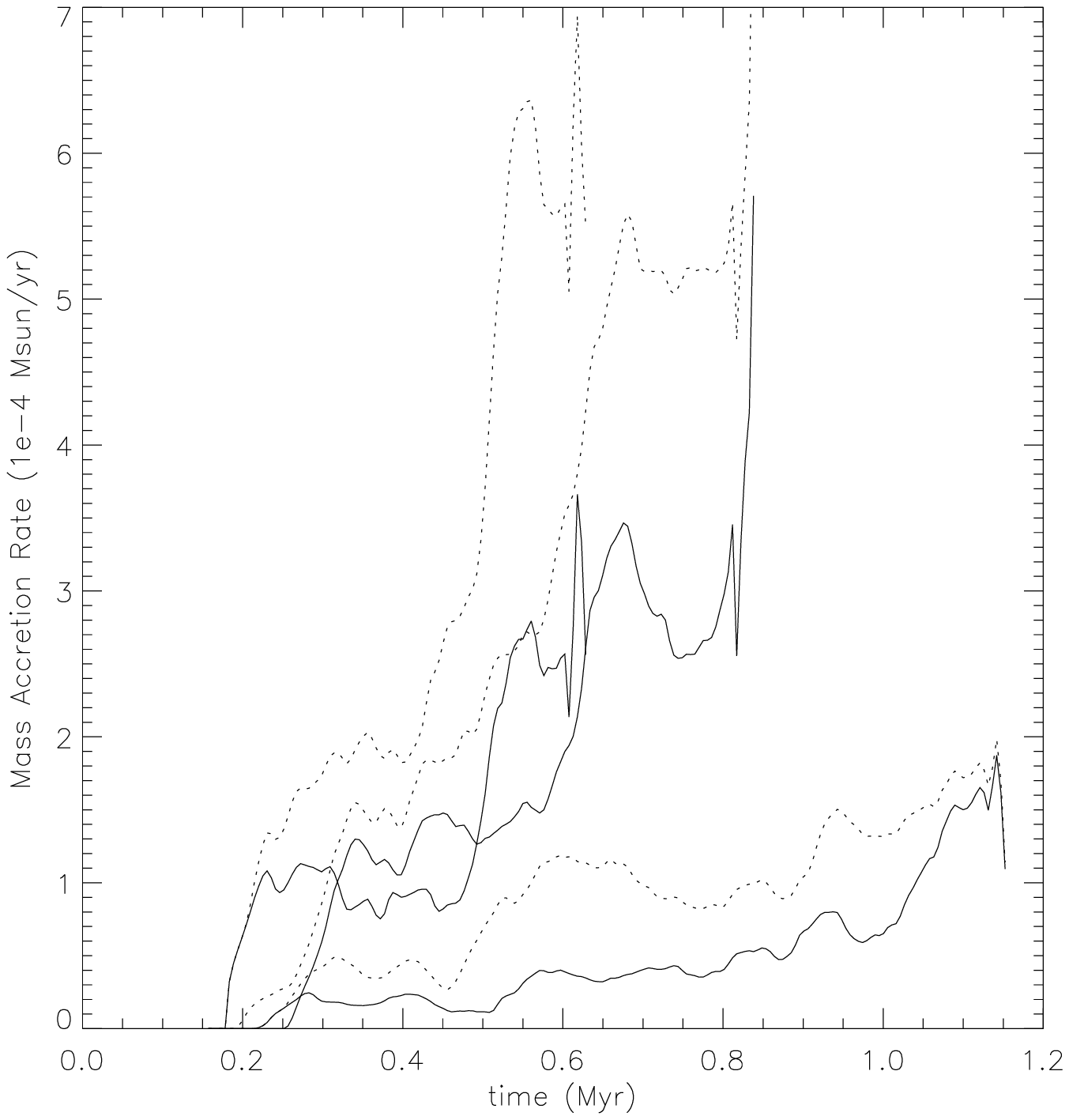}
\caption{Evolution of the total mass (top panel) and mass accretion
  rate (bottom panel) of all massive stars (dotted lines) and those 
  for the most massive object (solid lines). The curves from 
  upper-left to lower-right are for Model HD, MHD and WIND, respectively.} 
\label{massivestar}
\end{figure}

\begin{figure}
\epsscale{0.65}
\plotone{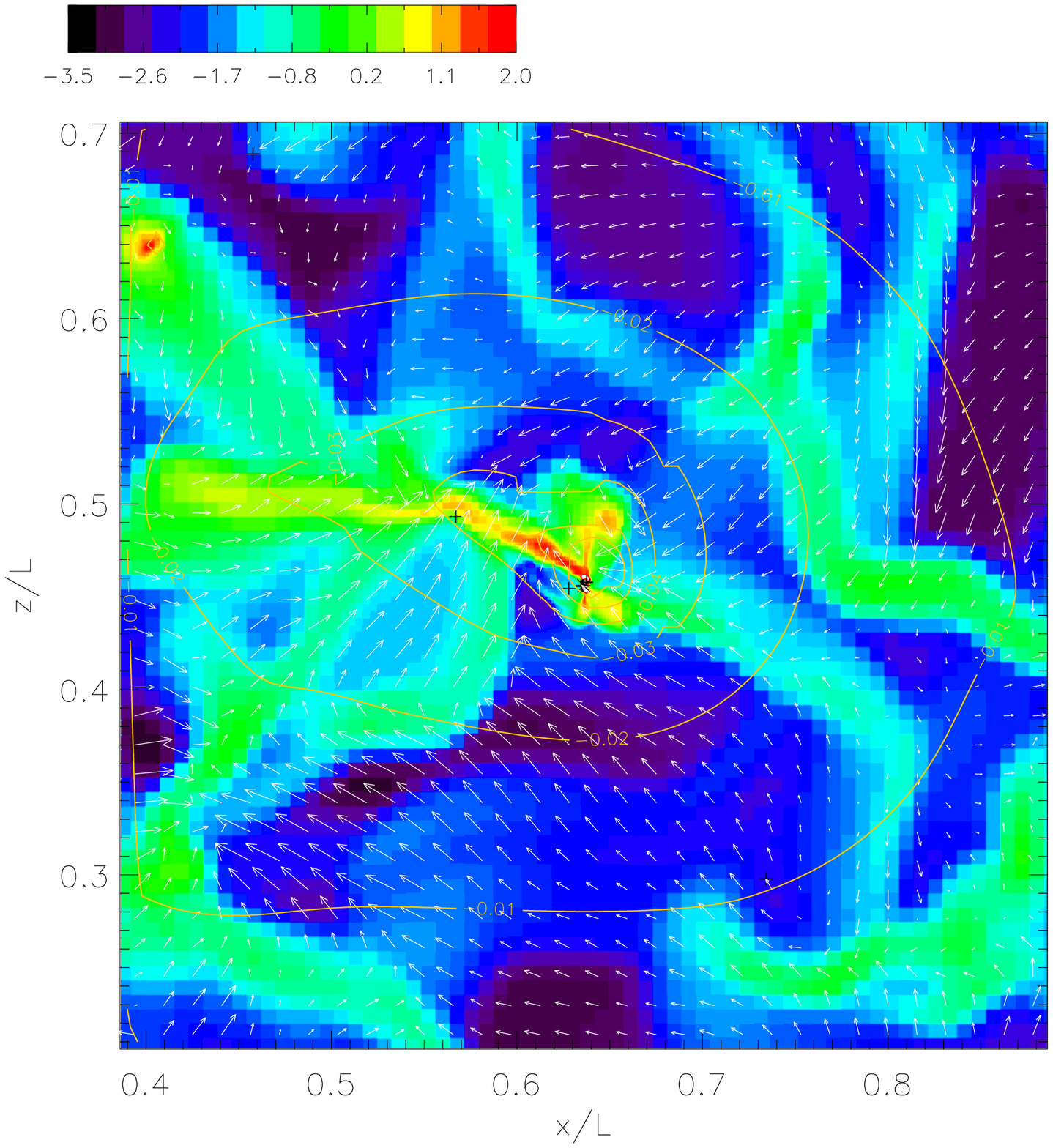}
\plotone{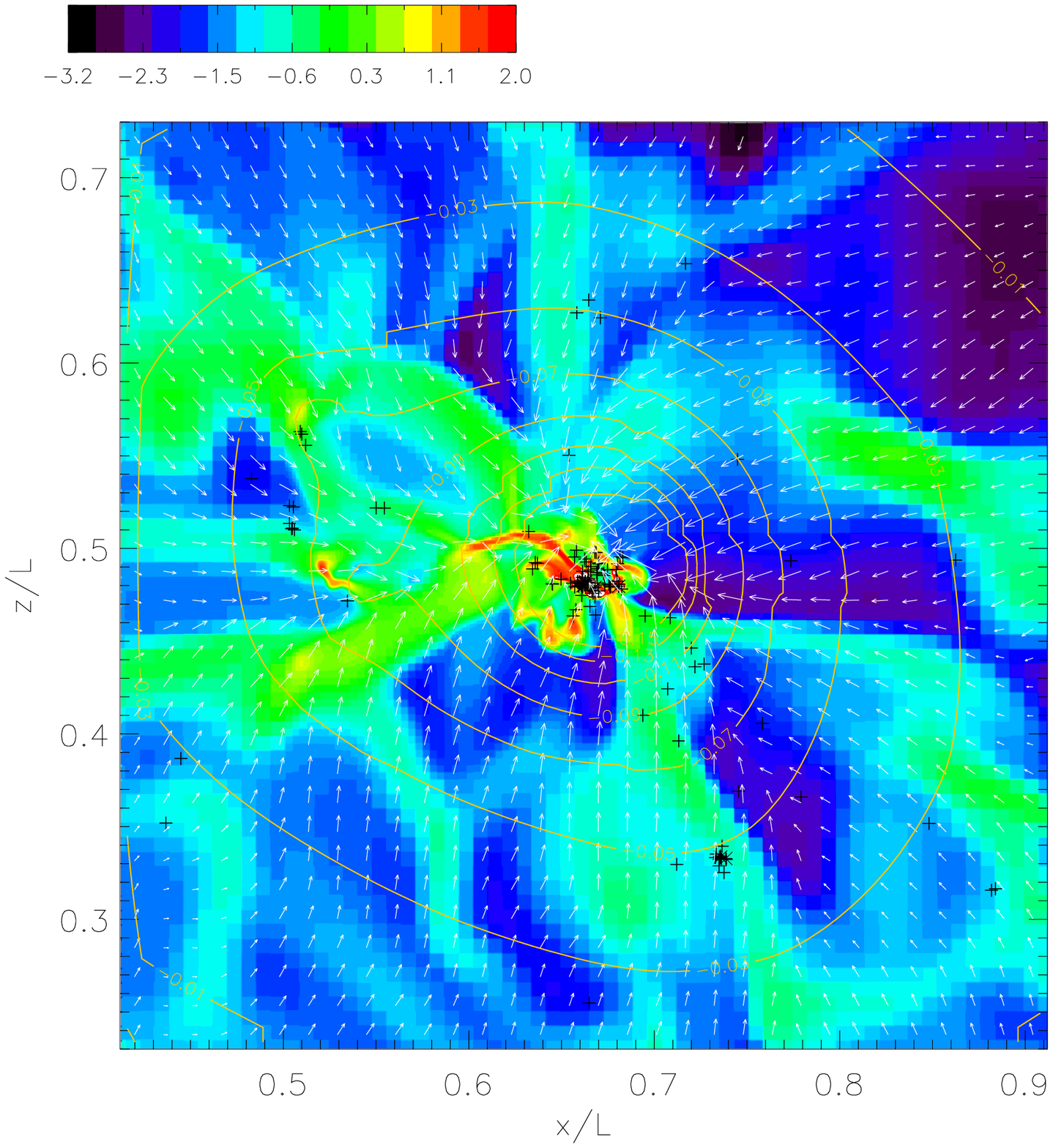}
\caption{Slice of the density through the most massive object for the HD 
model, with the velocity 
vectors (white) and contours of the gravitational potential overplotted. 
The first panel shows the early phase of rapid mass accretion for the 
most massive object through a dense filament at t=1.5. The second shows 
the global collapse at t=3. Stars are denoted by pluses.} 
\label{HDYslice}
\end{figure}

\begin{figure}
\epsscale{0.65}
\plotone{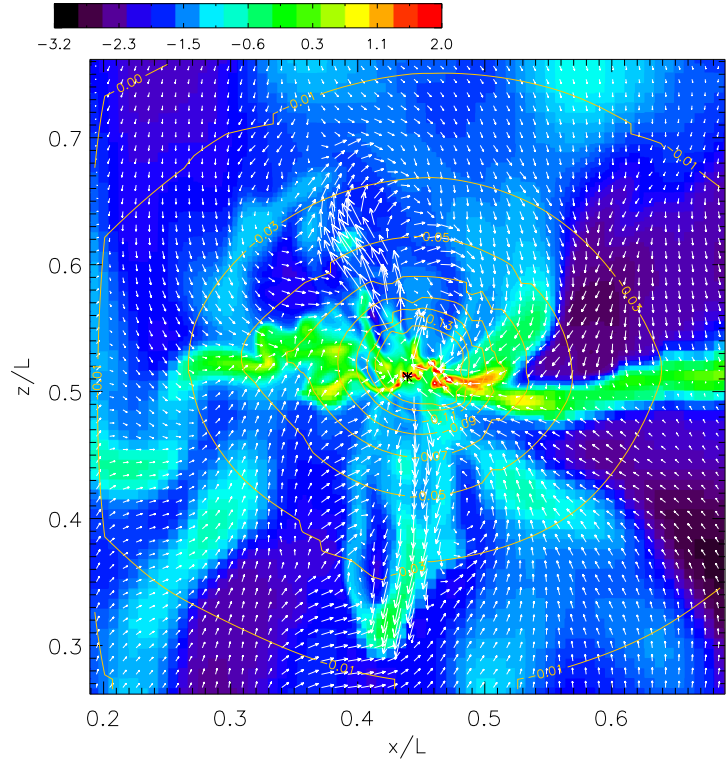}
\caption{Slice of the density through the most massive object (denoted
  by an asterisk), showing a large magnetic bubble driven from near 
  the central object at t=3.75 in the MHD Model.} 
\label{MHDYslice}
\end{figure}

\begin{figure}
\epsscale{1.0}
\plotone{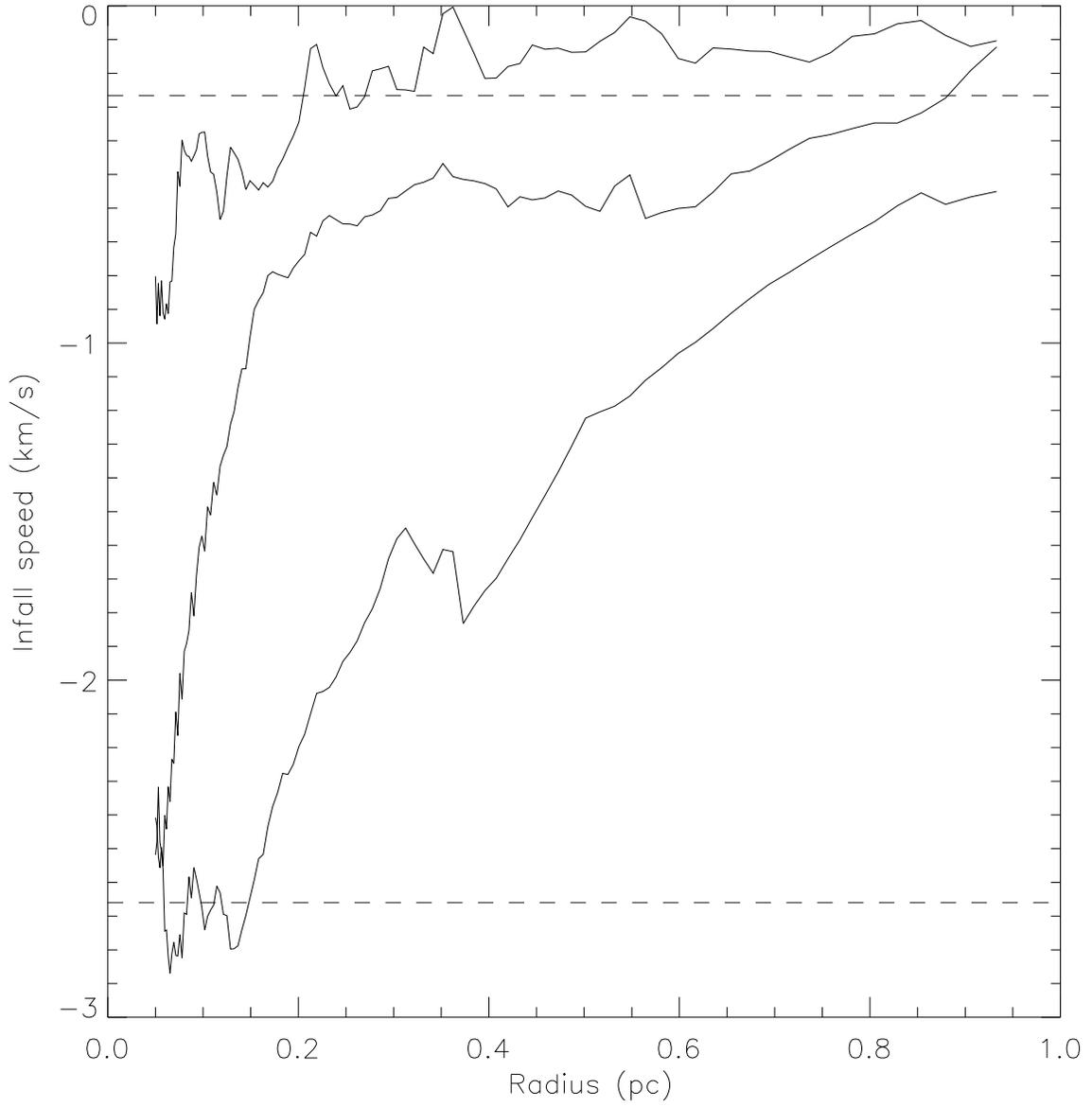}
\caption{Radially averaged, mass-weighted, infall speed towards the 
most massive object for Model HD (t=3, bottom curve), MHD (t=4, middle
curve) and WIND (t=5, top curve). The sound speed is plotted as dashed 
line for comparison.} 
\label{infall}
\end{figure}

\begin{figure}
\epsscale{0.65}
\plotone{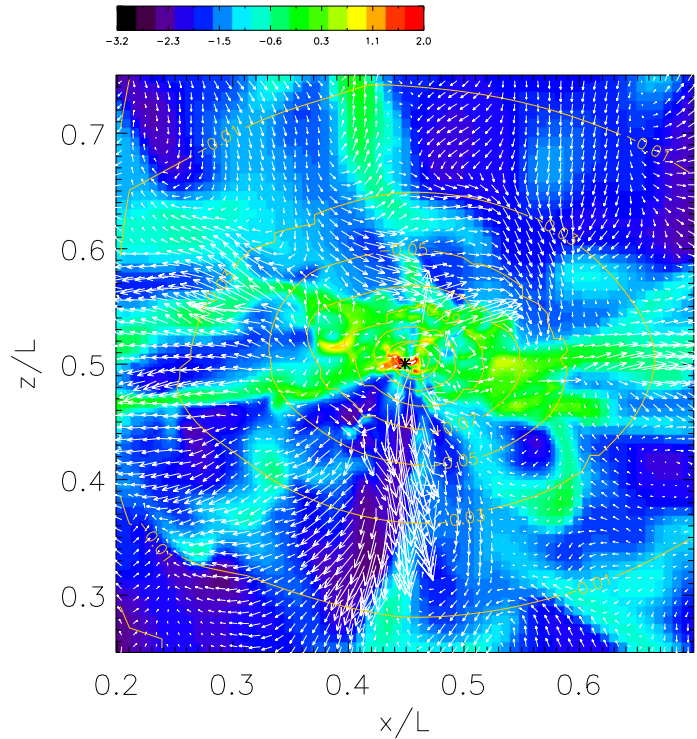}
\caption{Slice of the density through the most massive object (denoted
  by an asterisk) in the
  WIND model at $t=5$, showing a chaotic velocity field (denoted by 
  white arrows), as opposed to the ordered global collapse in the HD 
  case. } 
\label{WINDYslice}
\end{figure}

\begin{figure}
\epsscale{1.1}
\plotone{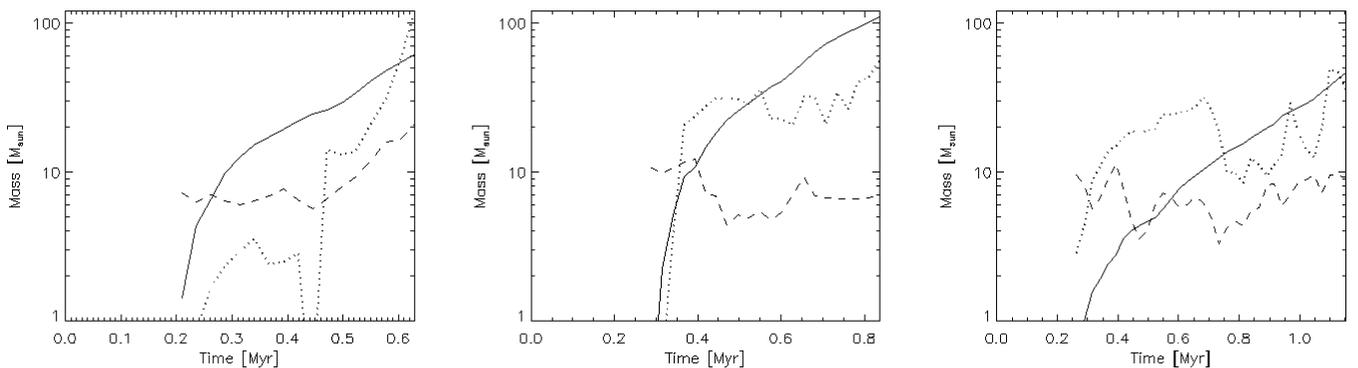}
\caption{Time evolution of the mass of the most massive object (solid), and
  the masses of the gas (dashed) and other stars (dotted) within a
  ``core'' (defined as a sphere of 0.1~pc in diameter centered  
 on the object) for the HD (left), MHD (middle) and WIND 
 (right) models. } 
\label{starcore}
\end{figure}

\end{document}